\documentclass[12pt]{article}

\usepackage{amsmath}
\usepackage{amssymb}
\usepackage[margin=1in]{geometry}
\usepackage{natbib}
\usepackage[capposition=top]{floatrow}
\usepackage{float}
\usepackage{subfigure}
\usepackage{eurosym}
\usepackage{bbm}
\usepackage{longtable}
\usepackage{booktabs}
\usepackage{adjustbox}
\usepackage{threeparttable}
\usepackage{enumitem}
\usepackage{graphicx}
\usepackage{xurl}
\usepackage{lscape}
\usepackage{changepage}
\usepackage{footmisc}
\usepackage{ragged2e}
\usepackage{setspace}
\usepackage[edges]{forest}
\usepackage{tikz}
\usepackage{tikz-qtree}
\usetikzlibrary{arrows.meta,shadows}
\usepackage[labelformat=parens,labelsep=quad,skip=3pt]{caption}
\usepackage{xcolor}
\usepackage{hyperref}
\hypersetup{
	colorlinks=true,
	linkcolor=blue,
	citecolor=blue,
	urlcolor=blue,
	pdftitle={Disaffection at Work}
}
\setlist[itemize]{noitemsep, topsep=2pt, parsep=0pt, partopsep=0pt}
\newcommand{\sym}[1]{\ifmmode^{#1}\else\(^{#1}\)\fi}

\begin{document}
	
	\thispagestyle{plain}
	
	\vspace*{0.75em}
	
	\begin{center}
		\begin{minipage}{0.82\textwidth}
			\centering
			{\Large \textbf{Disaffection at Work: Employee Responses to Job-Related Information}\textsuperscript{*}}
		\end{minipage}
		
		\vspace{1.4em}
		
		{\large
			Beatrice Braut\textsuperscript{\dag}
			\quad
			Mariele Macaluso\textsuperscript{\ddag}
			\quad
			Vincenzo Mollisi\textsuperscript{\ddag \S}
		}
		
		\vspace{1.1em}
		
		{\normalsize \today}
	\end{center}
	
	\vspace{2.25em}
	
	\begin{center}
		\begin{minipage}{0.74\textwidth}
			\begin{center}
				\textbf{Abstract}
			\end{center}
			
			\vspace{0.25em}
			
			\small
			\noindent
			Quiet quitting reflects a form of worker disaffection that operates along the intensive margin of labor supply rather than through job exit. We study how alternative workplace narratives shape workers' behavioral responses using a randomized survey experiment on a representative sample of employees in Italy and France. Respondents are exposed to empirically grounded moral framings of work emphasizing either social justice and collective rights or work organization and employment practices. We find that moral framings reallocate behavior across margins: justice-oriented narratives increase detachment while reducing passive disengagement, whereas organization-centered framings generate no systematic effects.
			
			\vspace{2.2em}
			
			
			\vspace{0.35em}
			
		\end{minipage}
	\end{center}
	
	\vfill
	
	\noindent\rule{\textwidth}{0.4pt}
	
	\vspace{0.35em}
	
	{\footnotesize
		\noindent
		\textsuperscript{*} This project was funded by the CLOSER Centre through the CLOSER–Qualtrics Call 2025. The study was preregistered at the AEA RCT Registry under ID AEARCTR-0015134. The preregistration plan is available at \url{https://www.socialscienceregistry.org/trials/15134}. Data and programs are available from the corresponding author upon request. 
		
		
		\noindent
		\textsuperscript{\dag}Department of Economics, University of Genova; \texttt{beatrice.braut@unige.it} \\
		\noindent
		\textsuperscript{\ddag}ESOMAS Department, University of Turin; \texttt{mariele.macaluso@unito.it}\\
		\noindent
		\textsuperscript{\S}DIGI Department, University of Palermo, and University of Turin; \texttt{vincenzo.mollisi@unipa.it}
	}
	
	\newpage
	
	\doublespacing
	\setcounter{page}{1}
	
	\section{Introduction}

This paper studies how workers respond to workplace dissatisfaction when exit is not the primary margin of adjustment. While recent discussions of the \emph{Great Resignation} have focused on job separations\footnote{The term \emph{Great Resignation} refers to the wave of voluntary job separations that followed the onset of the COVID-19 pandemic. Empirical evidence documents that quit rates during this period were higher than those observed in previous macroeconomic downturns, including the Great Recession \citep{AmanorBoadu2022}. One driver of this increase was workers' preference for remote and hybrid work arrangements: survey evidence shows that a substantial share of employees working from home at least part of the week would seek alternative employment rather than return full-time to business premises \citep{BarreroBloomDavis2021}.}, a growing share of post-pandemic labor market adjustment has taken place within ongoing employment relationships. In this context, workers may reduce effort, disengage from their tasks, or instead shift toward collective forms of voice. 
A growing public and academic debate has used the term \emph{quiet quitting} to describe workers who remain employed but mentally withdraw from their jobs, comply with basic job requirements, and
avoid additional responsibilities or initiative beyond what their role requires \citep{LeeParkShin2024, Bernuzzi2025}\footnote{The term became prominent in public debate in 2022 after spreading on TikTok, where it was used to describe a refusal to go beyond one's formal job duties and a rejection of burnout culture \citep{newport2022}. The phenomenon is often discussed alongside earlier anti-overwork movements such as the Chinese \emph{tang ping} (``lying flat'') movement, which emerged in 2021 as a response to intense productivity norms and long working hours \citep{cevero2025}.}. Rather than implying a break in the employment relationship, it captures an adjustment within the job, often associated with resistance to excessive workloads, work-related stress, and the erosion of work--life balance \citep{Galanis2023QuietQuitting}.

We examine how these alternative responses depend on the way workers interpret their employment relationship. The motivation for our analysis comes from two related strands of evidence. First, a large macroeconomic literature explains the rise in quits through standard labor market mechanisms, emphasizing efficient reallocation across firms. During periods of rapid expansion, stronger labor demand increases employer-to-employer transitions, while downward wage rigidity generates vacancy chains that amplify quits \citep{akerlof1988job, mercan2020jobs}. In particular, \citet{mercan2020jobs} show how job-to-job transitions propagate through vacancy chains, highlighting the central role of quits in reallocating workers across firms. Second, recent evidence shows that a substantial share of adjustment has instead occurred along the intensive margin. For instance, \citet{LeeParkShin2024} document that more than half of the post-2019 decline in aggregate hours worked in the United States is driven by reductions in hours rather than labor force exits, interpreting this decline in hours among employed workers as \emph{quiet quitting}. While the former margin is well understood, much less is known about how workers adjust behavior within existing matches.

A natural framework to organize these responses is provided by \citet{hirschman1970}, where dissatisfaction may lead to exit or voice, but may also manifest through changes in behavior within the employment relationship. Building on this framework, \citet{freeman1976, Freeman1978} shows that job satisfaction is a strong predictor of mobility, while also reflecting a dynamic evaluation of the job relative to outside opportunities. Importantly, dissatisfaction need not translate into exit, especially when adjustment costs are high or outside options are uncertain.

To address this gap, we implement a randomized survey experiment designed to isolate the causal effects of workplace narratives. Specifically, we randomly expose employed workers to alternative moral framings of workplace dissatisfaction and estimate their effects on stated behavioral responses. Respondents are exposed to wordclouds that make different dimensions of work-related discontent salient: social justice and collective rights, work organization and employment practices, or a combination of the two. We then measure responses across the main margins of adjustment emphasized by the exit--voice framework: disengagement at work, support for collective voice, quitting intentions, and preferences over wage--amenity trade-offs. The experiment is administered to a sample of 1,450 employed workers aged 20--45 in Northern Italy and in France, specifically the South and the Paris area. 

Italy and France offer a useful setting for studying adjustment to workplace dissatisfaction outside the exit margin. In both countries, post-pandemic evidence points to low workplace engagement and to the salience of intensive-margin responses within continuing employment relationships. In Italy, only 17 \% of workers are classified as fully engaged, and recent evidence documents a growing share of employees who remain in their jobs while reducing effort and initiative \citep{polimi2025}. In France, quiet quitting is similarly prominent in public and survey evidence, with 37 \% of workers reporting that they identify with limiting effort to the tasks formally required by their job \citep{ifop2022}. These patterns suggest that disaffection may be expressed not only through quits, but also through changes in effort, motivation, and workplace conduct.
The two countries are also informative because they combine relatively regulated labor markets with different traditions of workplace voice. At the same time, the two countries differ in the relative salience of voice and disengagement as margins of adjustment. France has a stronger tradition of collective representation and workplace mobilization, with labor relations shaped by state regulation, firm-level union representation, and collective bargaining over workplace outcomes \citep{AghionAlganCahuc2011, Breda2015}. Italy, by contrast, is characterized by high employment protection and lower job mobility, with labor-market reforms aimed at reducing dualism and increasing flexibility \citep{BoeriGaribaldi2019, BertoniChinettiNistico2023}. Comparing Italy and France therefore allows us to study whether similar dissatisfaction is expressed through different margins of adjustment, such as disengagement, voice, and quitting intentions, depending on the moral framing of the employment relationship.

A key challenge in studying such narratives is that explicit vignettes may induce experimenter demand effects or impose researcher-driven interpretations. To overcome this issue, we develop a novel framing strategy based on visually salient wordclouds constructed from large-scale textual data on discussions of work dissatisfaction. Rather than presenting respondents with explicit narratives, the wordclouds expose individuals to empirically grounded vocabularies of workplace concerns, allowing them to interpret the content through their own cognitive and moral frameworks. This approach activates distinct dimensions of the moral economy of work-related to social justice and collective rights, or to work organization and employment practices without imposing a predefined interpretation.

To capture how workers translate these interpretations into behavior, we measure a set of outcomes spanning multiple margins of adjustment. These include attitudinal measures of disengagement (such as detachment, lack of initiative, and reduced motivation), behavioral intentions related to workplace conduct, and forward-looking outcomes such as quitting intentions and support for collective voice. In addition, we construct preference-based measures capturing how workers trade off wages and non-monetary job attributes, using choices over wage–remote work bundles and willingness to accept wage reductions for additional work-from-home days, following recent work on willingness to pay for job amenities \citep{MasPallais2017, WiswallZafar2017, Maestas2023}.

Our results show that moral framings do not uniformly increase or decrease disengagement, but instead reallocate workers’ responses across behavioral margins. Exposure to justice-oriented narratives increases detachment by approximately 9 percentage points, while reducing passive forms of disengagement, such as lack of initiative, by 4 to 6 percentage points. By contrast, framings centered on work organization and employment practices generate no systematic effects across outcomes. We also find no evidence of short-run effects on quitting intentions or other forward-looking labor market behaviors.

\textbf{Related Literature.} This paper contributes to three strands of literature. First, we contribute to the literature on labor supply and job mobility. While quits are well understood as efficient reallocation across firms \citep{akerlof1988job, mercan2020jobs}, and the exit-voice tradition has long distinguished the margins along which dissatisfaction can operate \citep{hirschman1970, Freeman1978}, causal evidence on what determines \emph{which} margin is activated within ongoing matches remains scarce, as observed responses confound sorting, workplace conditions, and outside options. We provide experimental evidence on this allocation, showing that the framing of workplace dissatisfaction shifts responses across disengagement, voice, and exit intentions while holding the employment relationship constant.

Second, we contribute to the literature on non-wage job attributes and worker preferences. Because workers sort across firms, the value of amenities such as
flexibility and working conditions is typically identified through stated-preference designs over hypothetical job profiles \citep{Brown1980, MasPallais2017, WiswallZafar2017, HeNeumarkWeng2021, Maestas2023}, and a large body of evidence documents their role in shaping job choices and mobility \citep{Clark2001, sullivan2014search, Bloom2015, LeBarbanchon2021, NON2022102087, Dube2022, Folke2022, BarreroBloomDavis2023}. This literature varies the attributes of the job; we instead hold the job constant and vary only the frame through which workers evaluate it, showing that responses, including stated wage-amenity trade-offs, can shift even absent any change in wages, amenities, or contractual conditions.

Third, we relate to work on fairness, identity, and the moral dimensions of the employment relationship. A large economic literature shows that gift exchange, identity, peer comparisons, and perceived pay inequities affect effort, satisfaction,  productivity, quits, and search behavior \citep{Akerlof1982,AkerlofKranton2000, MasMoretti2009, CardMasMorettiSaez2012, OswaldProtoSgroi2015, BrezaKaurShamdasani2018, dube2019fairness}, while recent sociological evidence documents the moral repertoires through which workers justify disaffection \citep{varavallo2023moral}. The former typically varies workers' actual conditions or information about them, such as relative pay or peer productivity; the latter is descriptive. We bridge the two: taking the empirically observed moral repertoires of \citet{varavallo2023moral} as the basis of our treatments, we provide causal evidence that activating these frames, absent any information about the respondent's own job, reallocates workers' stated responses across behavioral margins.

The paper is organized as follows. Section~\ref{sec:Methods} presents the experimental design, including the construction of the wordcloud-based treatments and the main hypotheses. Section~\ref{sec:survey} describes the survey, the sample, the outcome measures, and the descriptive statistics. Section~\ref{sec:empirical} outlines the empirical strategy. Section~\ref{sec:Results} presents the main results and the heterogeneity analysis. Section~\ref{sec:Conclusions} concludes.

\section{Experimental design}
\label{sec:Methods}

Our experimental design builds on the \emph{moral economy} framework. The notion denotes the shared norms of fairness and mutual obligation that govern economic relations and whose violation triggers protest and collective action: it originates with \citet{polanyi1944great}, who emphasized the embedding of labor in social and moral structures, and \citet{thompson1971moral}, who showed how violations of community norms of fairness inspire collective action, and has been extended by \citet{sayer2007moral} through the concept of lay
normativity, people's everyday moral evaluations of economic life. The underlying idea is not foreign to economics: a long tradition shows that labor relations are governed by norms of reciprocity and fairness \citep{Akerlof1982, kahneman1986fairness} and by workers' identity \citep{AkerlofKranton2000}, and \citet{bowles2016moral} explicitly frames such social preferences as a moral economy operating alongside material incentives. Building on this tradition, \citet{varavallo2023moral} identify three dimensions shaping contemporary
narratives around work dissatisfaction, each rooted in one strand of the framework: \emph{Work and Employment} (organizational, Polanyian), \emph{Social Justice and Activism} (community-based, Thompsonian), and \emph{Health, Well-being, and Lifestyle} (individual, following Sayer). In the experiment, we vary the first two dimensions while holding the third constant.
The treatments activate these dimensions through visually salient keyword-based wordclouds.

\subsection{Construction of the treatments}
\label{sec:treatments}

The keywords displayed in the experimental wordclouds are drawn from the most frequent terms identified by \cite{varavallo2023moral} using BERTopic analysis of highly engaged posts from the \texttt{r/antiwork} subreddit. Rather than being researcher-generated, the keywords reflect empirically observed \emph{vocabularies of motive}, publicly shared justifications used in discussions about work dissatisfaction. We implement three treatment conditions and one control condition:
\begin{itemize}

\item \textbf{Activism (pure):} activates the \emph{Social Justice and Activism} dimension, emphasizing inequality, workers' rights, income disparities, and collective action.

\item \textbf{Work (pure):} activates the \emph{Work and Employment} dimension, focusing on organizational practices, wages, career trajectories, and management dynamics.

\item \textbf{Work + Activism (hybrid):} jointly activates the \emph{Work and Employment} and \emph{Social Justice and Activism} dimensions, combining organizational experience with distributive and rights-based concerns.

\item \textbf{Control (Amenities):} a neutral narrative centered on corporate amenities and symbolic workplace perks (e.g., office furniture, coffee machines, social events). This condition reflects typical corporate communication and serves as a baseline without explicit moral framing.

\end{itemize}

However, identifying the causal role of non-wage attributes is empirically challenging because workers sort across firms, as emphasized by the compensating differentials literature \citep{Brown1980, hwang1992, lavetti2023}.
 Recent studies have made significant progress on this issue by employing discrete choice experiments that present individuals with hypothetical job profiles varying in attributes and use stated preferences to infer their willingness to pay for these attributes \citep{eriksson2014, MasPallais2017, WiswallZafar2017, Maestas2023}.

In this paper, treatments are implemented through wordcloud visualizations (Figure~\ref{fig:wordcloud_ita}; the French version of the stimuli is reported in Appendix Figure~\ref{fig:wordcloud_fra}), where keyword size reflects their empirical frequency in the source data. Wordclouds activate moral frames in a non-narrative format, reducing experimenter demand effects and allowing respondents to interpret familiar discursive cues rather than explicit statements.

\begin{figure}[htbp]
\centering
\begin{tabular}{cc}

\subfigure[Control: Amenities]{
\includegraphics[width=0.45\linewidth]{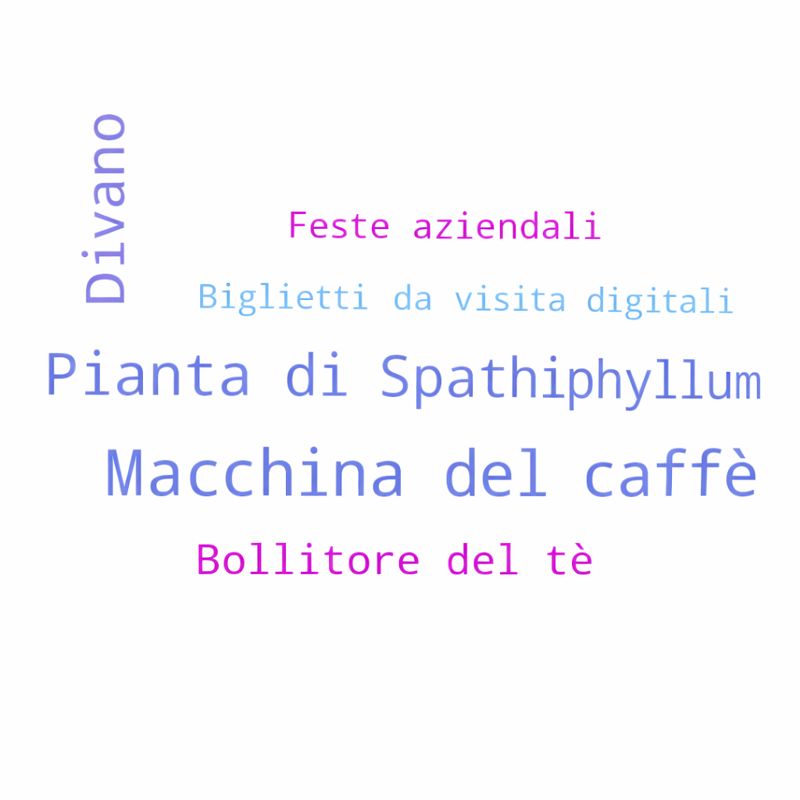}
}
&
\subfigure[T1: Activism]{
\includegraphics[width=0.45\linewidth]{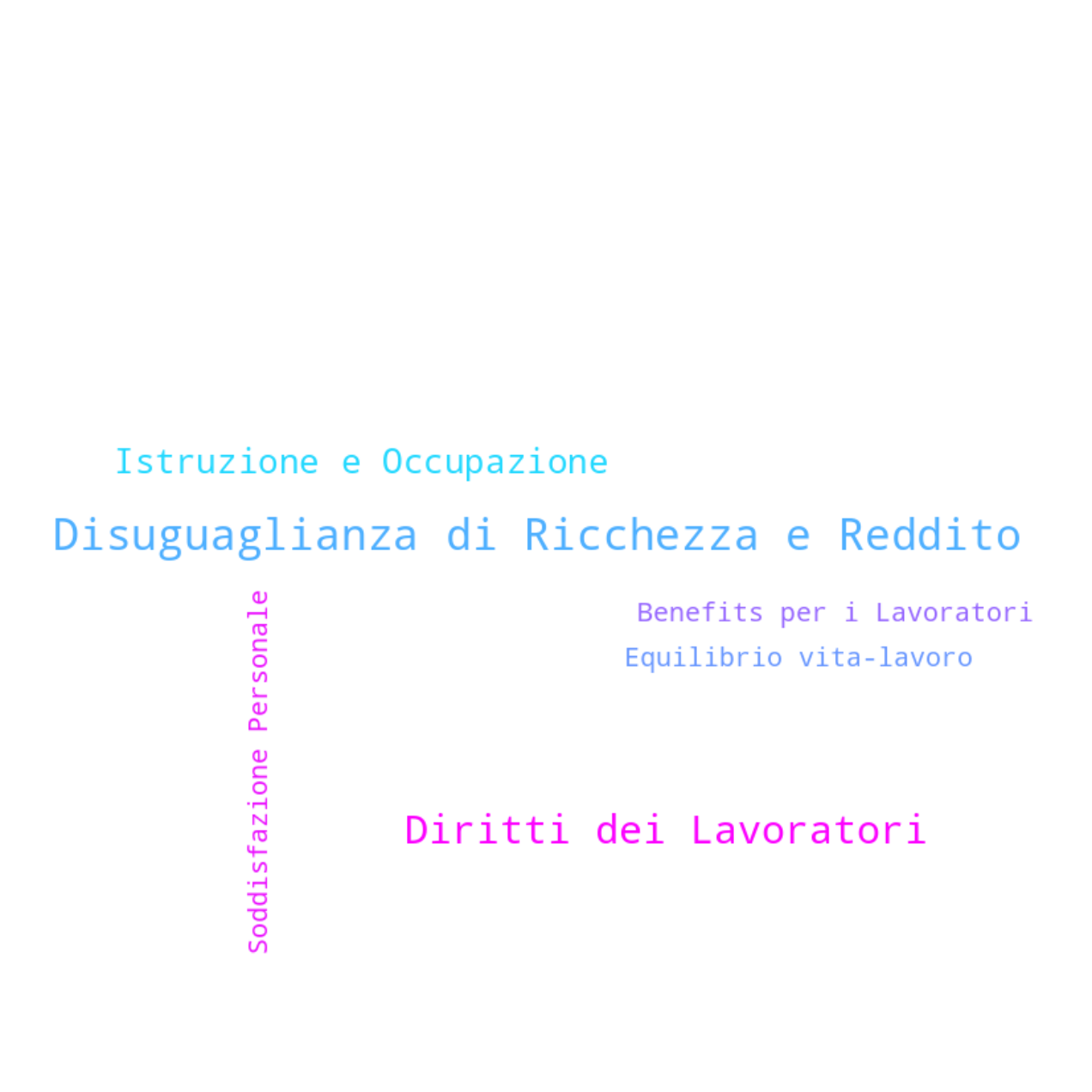}
}
\\[0.4cm]

\subfigure[T2: Work]{
\includegraphics[width=0.45\linewidth]{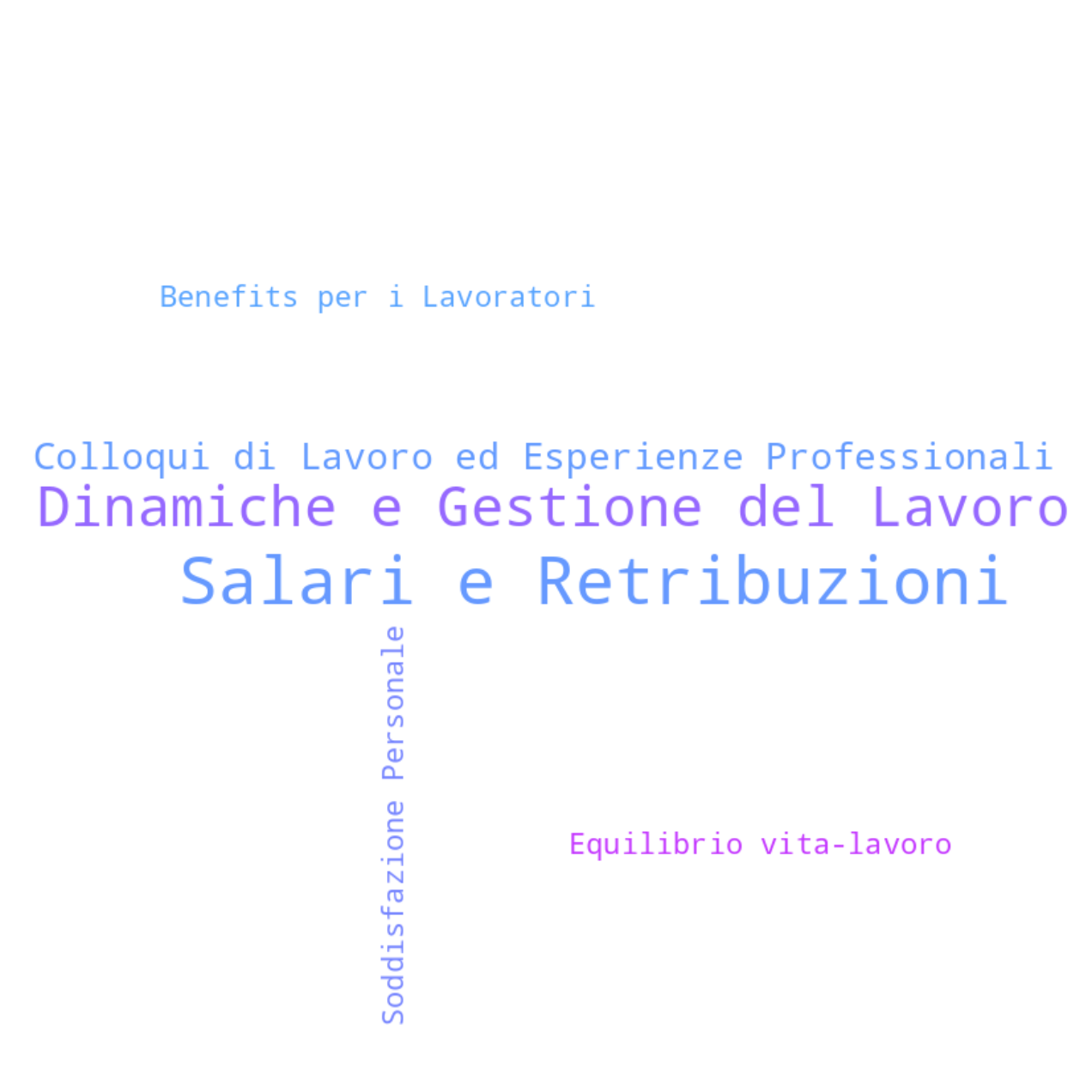}
}
&
\subfigure[T3: Work + Activism]{
\includegraphics[width=0.45\linewidth]{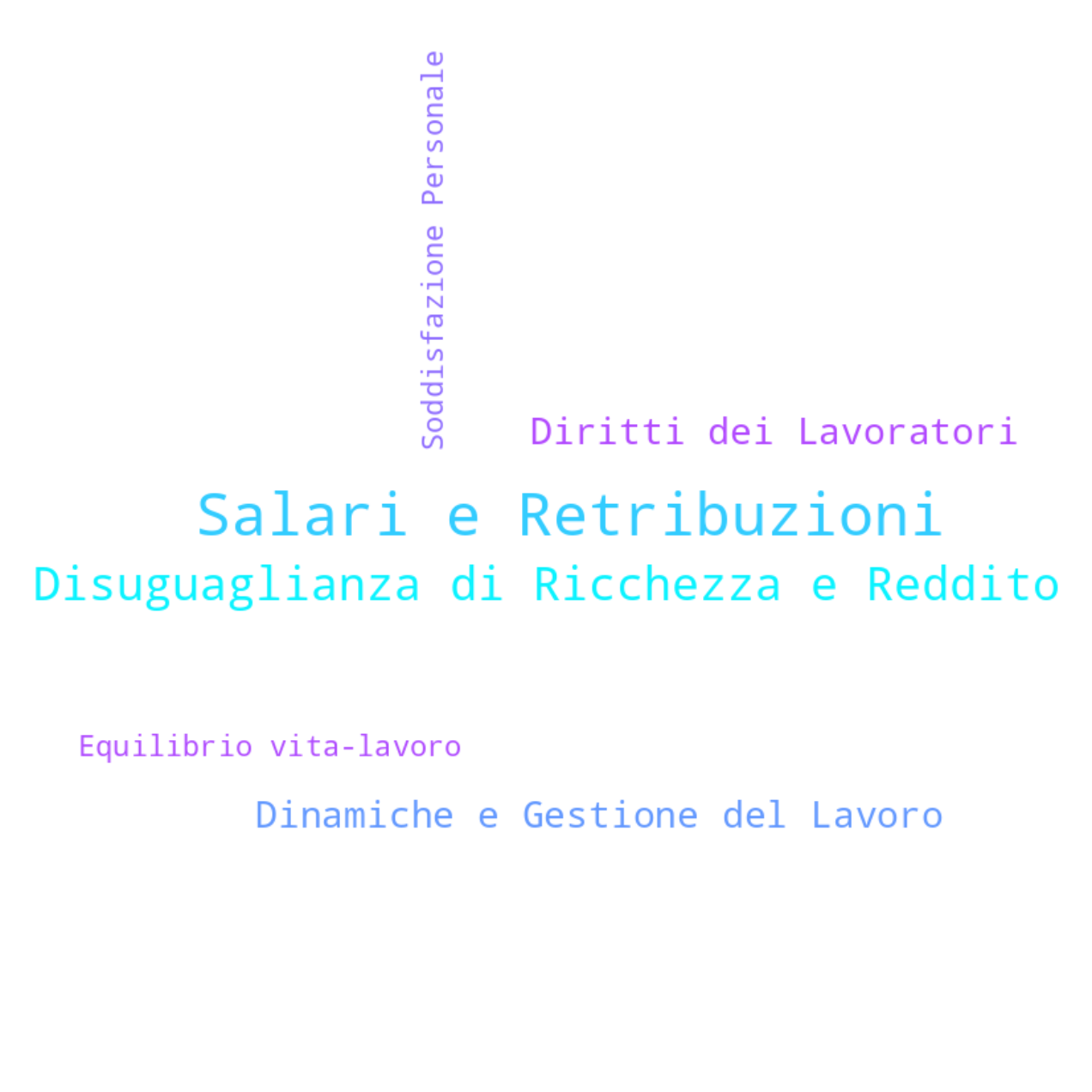}
}
\end{tabular}
\caption{Wordcloud-based experimental conditions (Italian).}
\label{fig:wordcloud_ita}
\vspace{0.4em}
\begin{minipage}{0.99\textwidth}
\footnotesize
\textit{Notes:} Wordcloud visualizations used as experimental stimuli (Italian version). 
Each panel represents one experimental condition. Word size reflects the relative frequency 
of keywords identified through BERTopic analysis of r/antiwork posts (\cite{varavallo2023moral}). To hold constant the Health, Well-being, and Lifestyle dimension the related terms are included symmetrically across treatments.
\end{minipage}
\end{figure}

\textbf{Exposure to the treatment.} The word-cloud stimulus is displayed on a dedicated survey page before the ranking task and outcome questions. The survey flow requires respondents to view the \textit{stimulus} before proceeding to the next page and prevents them from returning to previous pages thereafter.

Immediately after exposure, participants complete two ranking tasks. In the first task, respondents rank the keywords by the importance workers assign to each dimension. The task requires them to interpret the \textit{stimulus} and engage with the concepts highlighted in the visualization.

In the second task, respondents rank the same dimensions according to their own personal priorities. This second ranking provides a measure of the relative salience of different workplace dimensions for each respondent and is later used as a manipulation check to verify that the treatments successfully activate distinct moral economy channels.

\textbf{Multilingual implementation.} To ensure that treatment effects are not driven by translation artifacts, the same wordclouds were implemented in both Italian and French using conceptually equivalent translations of the original keywords. The visual structure and relative prominence of terms are held constant across languages, ensuring that treatment variation reflects differences in moral framing rather than linguistic salience.

\textbf{Neutralizing the Health, Well-being, and Lifestyle dimension.} While \cite{varavallo2023moral} identify health and lifestyle considerations as a third moral economy dimension, this channel primarily reflects individual circumstances, such as burnout or personal well-being that cannot be experimentally manipulated in a controlled manner. To avoid confounding treatment effects, each wordcloud includes the same small set of health-related keywords. As a result, variation across treatments is driven exclusively by differences in the \emph{Work and Employment} and \emph{Social Justice and Activism} dimensions.

\textbf{Manipulation check: Activation of moral economy channels.} A key requirement of our design is that the wordcloud-based treatments activate distinct moral economy channels rather than simply increasing general attention to work-related issues. To assess this, we exploit the second ranking task administered immediately after exposure to the experimental wordcloud.

In this task, respondents rank a set of workplace-related dimensions according to their personal importance at work. The list includes both \emph{treatment-specific} keywords (e.g., wages and management practices in the work condition; inequality and workers’ rights in the Activism condition) and a set of \emph{common} keywords (e.g., work–life balance, benefits, personal satisfaction) that are held constant across treatments. This structure allows us to construct a comparable measure of the relative salience of different moral economy dimensions.

For each respondent, we first invert the ranking scale so that higher values correspond to higher perceived importance. We then construct (i) a treatment-specific salience index as the average rank of keywords uniquely associated with the assigned moral channel, and (ii) a common-dimension index based on the shared keywords. The manipulation check measure is defined as the difference between the treatment-specific index and the common index. This gap captures the extent to which respondents prioritize channel-consistent dimensions relative to neutral dimensions that are common across treatments. Because the set of common keywords is identical across treatments, the resulting gap isolates the relative activation of the treatment-specific moral channel.

Appendix Table~\ref{tab:manipulation_check} reports the results. The two pure treatments generate clearly differentiated hierarchies of workplace priorities. In standardized terms, the relative salience gap differs by approximately 0.7 standard deviations between the Work and Activism conditions ($p < 0.001$), and the estimate remains stable when including the full set of socio-demographic and job-related controls.

These results indicate that the experimental stimuli successfully activate distinct evaluative channels. Rather than simply increasing overall attention to work-related issues, the treatments shift the relative importance assigned to workplace dimensions in a channel-consistent manner.

\subsection{Hypotheses}

\begin{figure}[H]
\centering
\begin{tikzpicture}[
    box/.style={rectangle, draw, rounded corners, align=center, minimum width=3.8cm, minimum height=1.1cm},
    arrow/.style={->, thick}
]


\node[box] (justice) at (0,3) {Social Justice \\ \& Activism};
\node[box] (work) at (0,1.7) {Work \& Employment};
\node[box] (health) at (0,0) {\footnotesize Health, Well-being \\ and Lifestyle \\ (held constant)};

\node[box] (t1) at (5,3.5) {T1: Activism};
\node[box] (t3) at (5,2.2) {T3: Work + Activism};
\node[box] (t2) at (5,0.8) {T2: Work};
\node[box] (c) at (5,-0.8) {\footnotesize Control: Amenities};

\node[box] (mech) at (10,1.5) {Mechanisms:\\
\footnotesize Detachment\\
Lack of initiative\\
Lack of motivation};


\draw[arrow] (work) -- (t2);
\draw[arrow] (justice) -- (t1);
\draw[arrow] (work) -- (t3);
\draw[arrow] (justice) -- (t3);

\draw[arrow] (t1) -- (mech);
\draw[arrow] (t2) -- (mech);
\draw[arrow] (t3) -- (mech);


\end{tikzpicture}
\caption{Conceptual framework: moral economy channels, treatments, mechanisms, and outcomes.}
\label{fig:conceptual_framework}
\end{figure}
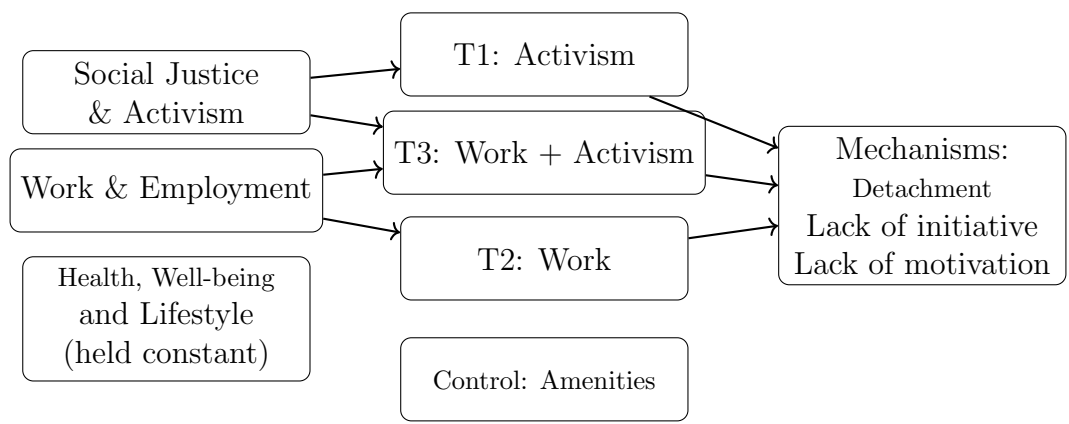

Figure~\ref{fig:conceptual_framework} summarizes the conceptual structure of the experiment. The treatments activate two distinct moral economy channels—\emph{Social Justice and Activism} and \emph{Work and Employment}—while holding the \emph{Health, Well-being, and Lifestyle} dimension constant. These framings may shape workers’ interpretations of workplace dissatisfaction and, through the mechanisms captured by the quiet-quitting scale, affect behavioral intentions related to disengagement, collective voice, and exit.

Building on this conceptual structure, we derive a set of hypotheses regarding the expected effects of each treatment on workers' behavioral intentions:

\textbf{H1 (Activism vs.\ Amenities).} Exposure to the Activism treatment activates a narrative centered on structural inequality and workers’ rights. This framing may shift workers’ interpretations of workplace dissatisfaction toward collective considerations, potentially affecting both disengagement and voice-related outcomes.

\textbf{H2 (Work vs.\ Amenities).} Exposure to the Work treatment emphasizes organizational practices and everyday workplace dynamics. By activating a frame focused on the employment relationship itself, this treatment may influence workers’ attitudes toward effort provision and engagement within the firm.

\textbf{H3 (Work + Activism).} The hybrid Work + Activism treatment jointly activates organizational and distributive dimensions of workplace dissatisfaction. By combining experiential and justice-oriented frames, this condition may generate responses that differ from those induced by the pure treatments.

Together, these hypotheses allow us to test whether alternative moral framings of workplace dissatisfaction shift workers’ responses across three possible psychological mechanisms: detachment, lack of initiative, and lack of motivation.

\section{Survey Design and Data}
\label{sec:survey}

The study is based on an original online survey experiment administered through Qualtrics Panels.

Qualtrics recruits respondents from large online panels and applies screening and quota procedures to approximate the population of interest, defined as full-time employed workers aged 20--45 residing in Northern Italy and in France, specifically the South and the Paris area. Soft quotas were implemented to ensure balance by gender (50--50) and age (32\% aged 20--28 and 68\% aged 29--45), as well as an approximately equal split between respondents in Italy and France.

The survey was fielded from March 11 to May 27, 2025. Data collection followed Qualtrics’ standard quality-control procedures. Responses exhibiting clear inconsistencies or implausible values, such as unrealistically low wages, invalid working hours or days, or completion times below pre-specified thresholds, were flagged and removed. These observations were replaced through additional collection waves to meet the target sample size. The final analytical sample consists of 1,450 completed responses.

Respondents were randomly assigned at the individual level to one of the experimental treatments described in Section~\ref{sec:treatments}. The survey followed a fixed sequence. It began with a set of pre-treatment socio-demographic questions (e.g., age, education, and household characteristics). Respondents were then exposed to the treatment stimulus and completed the ranking tasks and outcome measures. The final section collected additional job-related information about the respondent’s workplace and employment conditions. To minimize item non-response, all questions were mandatory and respondents could not skip items or return to previous pages. Partial responses were not recorded.

The survey includes several outcomes capturing workers’ attitudes and behavioral intentions. The core outcomes measure disengagement at work and are based on items from the Quiet Quitting Scale (QQS) developed by \citet{Galanis2023QuietQuitting}. These questions capture dimensions such as detachment, lack of initiative, and reduced motivation in the workplace.

In addition, we collect measures of broader labor-market responses, including the reported probability of leaving one’s job within the next 12 months, the likelihood of seeking support from trade unions, and the probability of reducing effort at work. The survey also includes preference-based questions related to remote work and compensation.

Full question wording and details on outcome construction are reported in Appendix~\ref{app:survey}.

\subsection{Outcomes and measurement}

We organize our outcome variables into two groups. The first group captures quiet quitting as a structured construct based on the Quiet Quitting Scale (QQS) developed by \citet{Galanis2023QuietQuitting}. The second group includes additional outcomes that capture broader labor-market responses, including exit intentions, collective voice, effort reduction, and preferences over work–life trade-offs.


\textbf{Quiet quitting outcomes.} In their original contribution, \citet{Galanis2023QuietQuitting} propose a scale composed of nine items loading on three latent dimensions: detachment, lack of initiative, and lack of motivation.

For the purposes of the present experiment, we construct a balanced set of six agreement-based items by selecting two items for each latent dimension. This choice ensures that each dimension contributes symmetrically to the analysis while maintaining comparability across constructs. All selected items follow the original wording, with respondents indicating their agreement on a four-point Likert scale ranging from “strongly disagree” to “strongly agree.” Items are coded so that higher values consistently indicate higher levels of quiet quitting.

The selection of items follows three criteria. First, we exclude questions framed in terms of behavioral frequency (“How often”), which rely on a different cognitive frame and are analyzed separately. Second, we exclude the item referring to delegating one’s tasks to colleagues, as responses depend on team composition, which is not controlled in the experimental design. Third, among two closely related items capturing silence at work, we retain the item referring to fear of being assigned additional tasks and drop the item referring to the belief that working conditions will not change, in order to avoid redundancy.

In addition, the first question in the block asks respondents whether they agree with the statement “I give my best at work.” This item is not included in the quiet quitting index but serves as a benchmark measure of overall engagement and helps mitigate potential anchoring effects at the beginning of the question block.

Table~\ref{tab:qqs_items} reports the mapping between the original QQS items and the subset retained in the present study.

\begin{table}[htbp]\centering
\caption{Mapping of original QQS items to latent factors and inclusion in the analysis}
\label{tab:qqs_items}
\begin{tabular}{p{9.5cm} l c}
\hline\hline
\textbf{Original item (\cite{Galanis2023QuietQuitting})} & \textbf{Factor} & \textbf{Included} \\
\hline
I find motives in my job. & Lack of motivation & Yes \\
I feel inspired when I work. & Lack of motivation & Yes \\
I do the basic or minimum amount of work without going above and beyond. & Detachment & Yes \\
If a colleague can do some of my work, then I let him/her do it. & Detachment & No \\
I don’t express opinions and ideas about my work because I am afraid that the manager assigns me more tasks. & Lack of initiative & Yes \\
I don’t express opinions and ideas about my work because I think that working conditions are not going to change. & Lack of initiative & No \\
I take as many breaks as I can. & Detachment & Yes \\
I often take initiative at work. & Lack of initiative & Yes \\
How often do you pretend to be working in order to avoid another task? & Detachment & No \\
\hline\hline
\end{tabular}

\begin{minipage}{0.99\textwidth}
\footnotesize
\textit{Notes:} The table reports the original items proposed by \citet{Galanis2023QuietQuitting} and their assignment to latent factors based on the exploratory factor analysis with oblique rotation (promax). The last column indicates whether the item is retained in the present analysis.
\end{minipage}
\end{table}


\textbf{Behavioral frequency outcomes.} Beyond agreement-based statements, the original QQS includes behavioral questions framed as “How often.” In our design, these items are analyzed separately rather than aggregated with the agreement-based scale, in order to avoid combining attitudinal responses with reported behavioral frequencies.

We retain most of the original frequency-based items, excluding the question on answering work-related calls after one’s shift because it conceptually overlaps with the item on answering messages on days off. The remaining questions capture behaviors such as helping colleagues complete tasks, responding to work-related messages during days off, arriving late or leaving early, taking sick leave despite being able to work, and pretending to work to avoid additional tasks. Table~\ref{tab:freq_items} summarizes the frequency-based items included in the analysis.

\begin{table}[htbp]\centering
\caption{Frequency-based items from the Quiet Quitting Scale}
\label{tab:freq_items}
\resizebox{\textwidth}{!}{%
\begin{tabular}{lcc}
\hline\hline
\textbf{Original item (Galanis et al.)} & \textbf{Status in (Galanis et al.)} & \textbf{Included in paper} \\
\hline
How often do you \textbf{help your colleagues} when you have completed your own tasks? & Excluded & Yes \\
How often do you answer phone calls/messages from your work after your work shift? & Excluded & No \\
How often do you \textbf{answer} phone calls/messages from your work \textbf{on a day off}? & Excluded & Yes \\
How often do you \textbf{go to work later and/or leave work early}? & Excluded & Yes \\
How often do you \textbf{take sick leave} even though you can work? & Excluded & Yes \\
How often do you \textbf{pretend to be working} in order to avoid another task? & Retained & Yes \\
\hline\hline
\end{tabular}}

\begin{minipage}{0.99\textwidth}
\footnotesize
\textit{Notes:} The table reports the frequency-based items originally proposed by \citet{Galanis2023QuietQuitting} and indicates whether they are included in the present analysis.
\end{minipage}
\end{table}

\subsection{Additional outcomes}
The second group of outcomes captures broader labor-market responses related to quiet quitting.
The first set refers to respondents’ current employment situation. In particular, respondents report subjective probabilities (on a 0-100 scale) of leaving their job within the next twelve months, seeking support from trade unions, and reducing effort at work. These measures capture potential behavioral responses within the respondent’s actual workplace and allow us to distinguish quiet quitting from classic exit and collective voice.

The survey also includes outcomes based on hypothetical scenarios designed to include preferences over workplace flexibility and monetary compensation. Respondents are presented with a policy scenario in which employers must compensate workers for work that cannot be performed remotely. They are then asked to choose among alternative combinations of remote work and salary increases, ranging from full remote work with no pay increase to no remote work with a substantial salary increase.
In addition, we asked respondents for their willingness to give up a share of their net salary to work one additional day at home. Together, these measures provide complementary evidence on how workers value remote work relative to monetary compensation.

These outcomes allow us to distinguish between the behavioral intentions within the current employment relationships of the respondents and the preferences expressed in hypothetical scenarios involving alternative compensation and workplace arrangements.

\subsection{Descriptive statistics}

This section presents descriptive statistics for the main outcomes, workplace behaviors, and workers’ actions and preferences. It also reports summary statistics for treatment assignment and for individual and job characteristics.

Table~\ref{tab:summary_statistics_outcomes} presents summary statistics for the main outcomes, workplace behaviors, stated actions and preferences, and treatment assignment. 

Panel A reports outcomes measured on four-point Likert scales. Overall, response indicates moderate levels of detachment and lack of initiative, with average values around 2.1--2.2 for doing only the minimum amount of work, taking many breaks, and not sharing ideas for fear of additional tasks. At the same time, the mean value for taking initiative at work is 3.00, suggesting variation across dimensions of workplace motivation and initiative.
Panel B reports workplace behaviors measured on five-point frequency scales. It shows that prosocial workplace behavior is relatively frequent: helping colleagues has the highest mean, equal to 3.59 on a five-point scale. By contrast, less cooperative behaviors such as arriving late or leaving early, calling in sick, and pretending to work have lower average values, close to or below 2. 
Panel C shows that respondents report average subjective probabilities of about 25--28 percent for quitting their job, seeking union help, or reducing work effort. 
Finally, Panel D confirms that treatment assignment is approximately balanced across the four experimental groups, with each group representing about one quarter of the sample.

\begin{table}[h!]
\centering
\caption{Summary Statistics: Outcomes}
\label{tab:summary_statistics_outcomes}
\footnotesize
\setlength{\tabcolsep}{3pt}
\renewcommand{\arraystretch}{0.82}

\begin{adjustbox}{max totalsize={\textwidth}{0.82\textheight},center}
\begin{threeparttable}
\begin{tabular}{p{9.2cm}ccccc}
\toprule
                    & \textbf{N} & \textbf{Mean} & \textbf{SD} & \textbf{Min} & \textbf{Max} \\
\midrule

\multicolumn{6}{l}{\textit{Panel A. Outcomes}} \\
Basic or minimum amount of work  (Detachment I)                 & 1433 & 2.20 & 0.93 & 1.00 & 4.00 \\
Take as many breaks as possible (Detachment II)                 & 1431 & 2.13 & 0.89 & 1.00 & 4.00 \\
Do not share ideas (fear of extra tasks) (Lack of Initiative I) & 1431 & 2.12 & 0.90 & 1.00 & 4.00 \\
Take initiative at work  (Lack of Initiative II)                & 1431 & 3.00 & 0.74 & 1.00 & 4.00 \\
I find motives in my job (Lack of Motivation I)                 & 1435 & 1.84 & 0.89 & 1.00 & 4.00 \\
I feel inspired when I work (Lack of Motivation II)             & 1436 & 2.86 & 0.82 & 1.00 & 4.00 \\

\addlinespace[0.5em]
\multicolumn{6}{l}{\textit{Panel B. Workplace behaviors}} \\
Help colleagues                          & 1442 & 3.59 & 1.00 & 1.00 & 5.00 \\
Answer on days off                       & 1442 & 2.99 & 1.27 & 1.00 & 5.00 \\
Arrive late / leave early                & 1442 & 2.08 & 1.20 & 1.00 & 5.00 \\
Call in sick                             & 1442 & 1.86 & 1.17 & 1.00 & 5.00 \\
Pretend to be working                    & 1442 & 2.05 & 1.23 & 1.00 & 5.00 \\

\addlinespace[0.5em]
\multicolumn{6}{l}{\textit{Panel C. Actions and preferences}} \\
Quit job within 12 months                & 1442 & 27.86 & 31.23 & 0.00 & 100.00 \\
Seek union help                          & 1442 & 24.76 & 28.10 & 0.00 & 100.00 \\
Reduce work effort                       & 1442 & 26.12 & 28.59 & 0.00 & 100.00 \\
Preferred compensation package           & 1442 & 3.14 & 1.95 & 1.00 & 6.00 \\
Accepted pay cut for remote work         & 1442 & 5.08 & 2.14 & 1.00 & 7.00 \\

\addlinespace[0.5em]
\multicolumn{6}{l}{\textit{Panel D. Treatment assignment}} \\
Activism treatment                       & 1419 & 0.25 & 0.43 & 0.00 & 1.00 \\
Work-active treatment                    & 1419 & 0.24 & 0.43 & 0.00 & 1.00 \\
Work treatment                           & 1419 & 0.26 & 0.44 & 0.00 & 1.00 \\
Control group                            & 1419 & 0.25 & 0.44 & 0.00 & 1.00 \\

\bottomrule
\end{tabular}

\begin{tablenotes}[flushleft]
\scriptsize
\item This table reports summary statistics for the main study variables. Panel A variables are measured on a four-point Likert scale; Panel B variables on a five-point frequency scale. Variables in Panel C include subjective probabilities ranging from 0 to 100 and ordered preference measures. Summary statistics are computed on the restricted sample of respondents with survey duration between 194 and 2350 seconds.
\end{tablenotes}
\end{threeparttable}
\end{adjustbox}
\end{table}

Table~\ref{tab:summary_statistics_demographics} describes the individual, demographic, and job characteristics of the estimation sample. The sample is broadly balanced by gender, with women representing about half of respondents. It also includes respondents with diverse household characteristics: 61 \% report living with a partner, 49 \% have children, and 67 \% reside in urban areas.
Respondents display relatively high labor-market attachment. On average, they have been in their current job for 8.27 years and have accumulated 13.69 years of total work experience. The mean commute time is 36 minutes, but the large dispersion in this variable points to the presence of extreme values.
The table also shows substantial variation in job characteristics. Twenty-six percent of respondents work remotely, 43 \% are employed in large firms, 37 \% work in the public sector, and 17 \% are part-time workers. The occupation and sector variables further indicate that the sample spans different types of jobs and industries\footnote{We aggregate occupations into three groups: high-skill (managers and senior officials; professionals), medium-skill (associate professionals; technicians; clerical occupations), and low-skill (trades; armed forces; plant and machine operators and assemblers; elementary occupations). We similarly aggregate industries into three broad sectors: primary (agriculture, forestry and fishing; mining and quarrying), secondary (manufacturing; electricity, gas, steam and air conditioning; water supply, sewerage, waste management and remediation; construction), and tertiary (wholesale and retail trade; transportation and storage; accommodation and food services; information and communication; financial and insurance activities; real estate; professional, scientific and technical activities; administrative and support services; public administration and defense; education; human health and social work; arts, entertainment and recreation; other services; household activities; extraterritorial organizations and bodies).}. The sample exhibits substantial heterogeneity in both socio-demographic characteristics and employment conditions.

\begin{table}[h!]
\centering
\caption{Demographic and job characteristics}
\label{tab:summary_statistics_demographics}
\footnotesize
\setlength{\tabcolsep}{3pt}
\renewcommand{\arraystretch}{0.82}

\begin{adjustbox}{max totalsize={\textwidth}{0.82\textheight},center}
\begin{threeparttable}
\begin{tabular}{p{5.5cm}ccccc}
\toprule
                    & \textbf{N} & \textbf{Mean} & \textbf{SD} & \textbf{Min} & \textbf{Max} \\
\midrule

Female                                   & 1442 & 0.50 & 0.50 & 0.00 & 1.00 \\
Couple                                   & 1441 & 0.61 & 0.49 & 0.00 & 1.00 \\
Urban                                    & 1442 & 0.67 & 0.47 & 0.00 & 1.00 \\
Children                                 & 1439 & 0.49 & 0.50 & 0.00 & 1.00 \\
Age group                                & 1442 & 3.25 & 1.39 & 1.00 & 5.00 \\
Years in current job                     & 1442 & 8.27 & 7.41 & -1.00 & 45.00 \\
Years of work experience                 & 1442 & 13.69 & 8.49 & 0.00 & 45.00 \\
Commute time (minutes)                   & 1442 & 36.09 & 292.67 & 0.00 & 10000.00 \\
Income category                          & 1393 & 1.96 & 0.83 & 1.00 & 5.00 \\
Remote work                              & 1442 & 0.26 & 0.44 & 0.00 & 1.00 \\
Large firm                               & 1442 & 0.43 & 0.50 & 0.00 & 1.00 \\
Public sector                            & 1442 & 0.37 & 0.48 & 0.00 & 1.00 \\
Part-time                                & 1442 & 0.17 & 0.38 & 0.00 & 1.00 \\
Occupation group                         & 1442 & 2.11 & 0.65 & 1.00 & 3.00 \\
Sector                                   & 1442 & 2.78 & 0.47 & 1.00 & 3.00 \\
French                                   & 1442 & 0.50 & 0.50 & 0.00 & 1.00 \\

\bottomrule
\end{tabular}

\begin{tablenotes}[flushleft]
\scriptsize
\item Summary statistics are computed on the restricted sample of respondents with survey duration between 194 and 2350 seconds. Years in current job and years of work experience are derived from reported starting years. The minimum of years in current job reflects one observation with reported starting year equal to 2026.
\end{tablenotes}
\end{threeparttable}
\end{adjustbox}
\end{table}

\begin{figure}[h!]
\centering
\includegraphics[width=\textwidth]{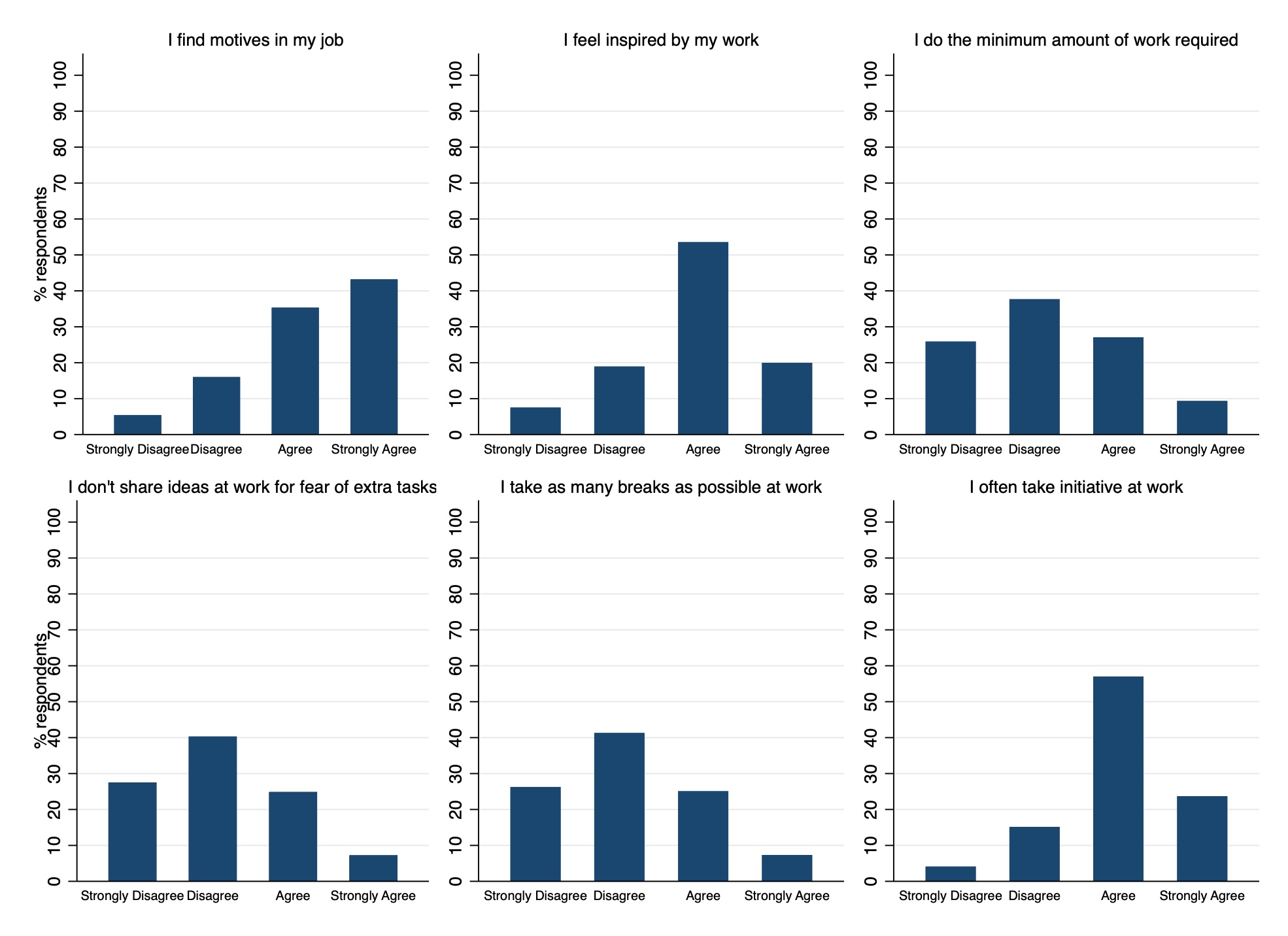}
\caption{Distribution of responses across quiet quitting dimensions}
\label{fig:vignette_2_7_combined_clean}
\begin{minipage}{\textwidth}
\footnotesize
\textit{Notes:} The figure reports the distribution of responses to agreement-based items capturing dimensions of quiet quitting. Responses are measured on a four-point Likert scale ranging from ``strongly disagree'' to ``strongly agree.'' Bars represent the percentage of respondents in each category. The sample includes respondents with survey completion times between 194 and 2350 seconds.
\end{minipage}
\end{figure}

Figure~\ref{fig:vignette_2_7_combined_clean} reports the distribution of responses to the agreement-based items capturing key dimensions of quiet quitting, measured on four-point Likert scales.

The results suggest that a lack of motivation at work is not widespread, while most report feeling inspired, pointing to substantial baseline attachment. At the same time, there is evidence of partial disengagement along specific margins. A significant share of respondents agree that they perform only the minimum amount of work required, suggesting that reductions in discretionary effort are relatively common. Responses to items capturing initiative display a similar pattern, with a non-trivial fraction of individuals reporting reluctance to express ideas due to the risk of being assigned additional tasks. By contrast, more passive forms of disengagement, such as taking breaks, are concentrated in intermediate categories, with relatively little mass at the extremes.

\begin{figure}[htbp]
\centering
\includegraphics[width=\textwidth]{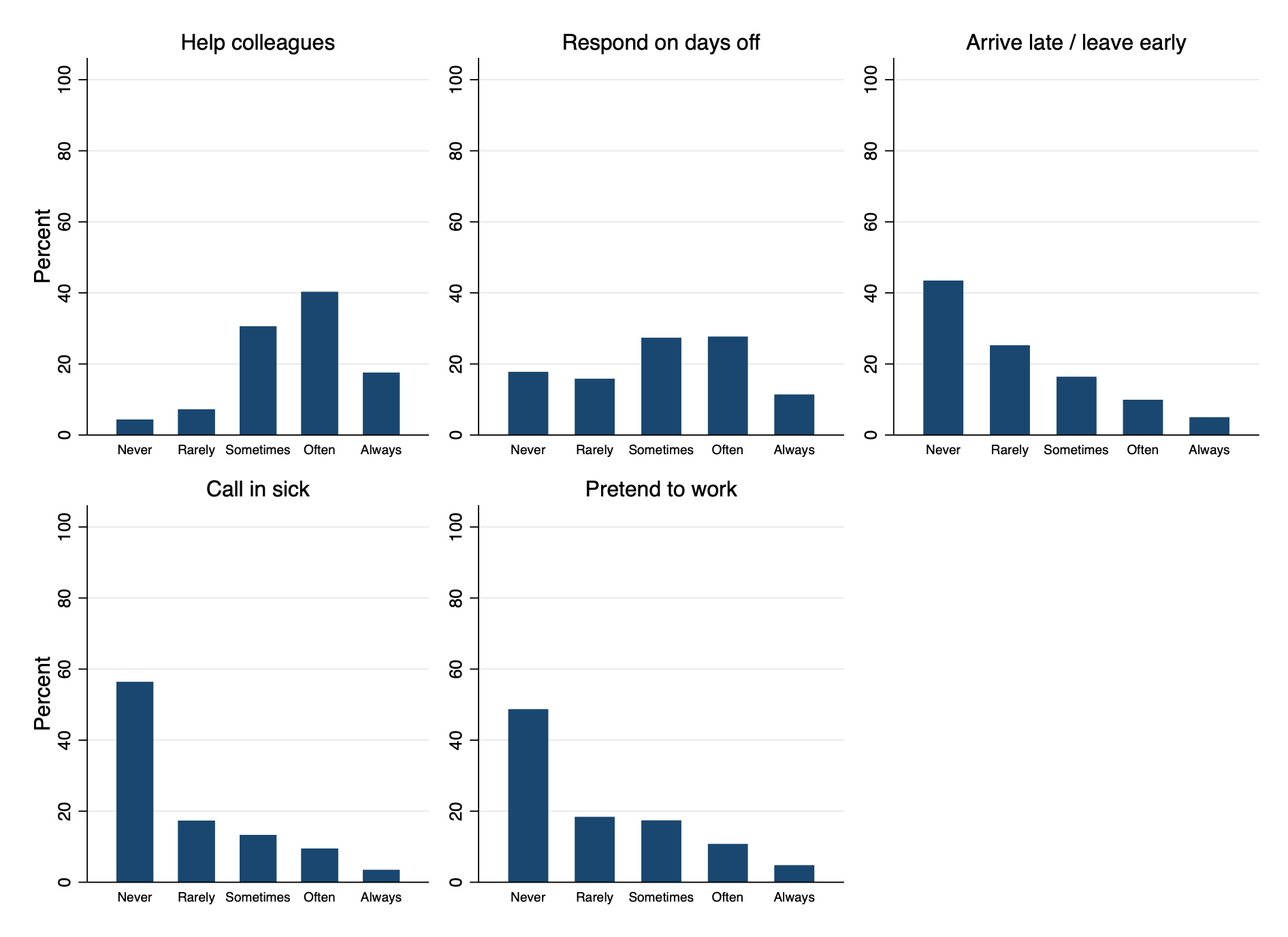}
 \caption{Distribution of responses to quiet quitting frequency items}
\label{fig:vignette_freq_combined}
\begin{minipage}{\textwidth}
\footnotesize
\textit{Notes:} The figure reports the distribution of responses to frequency-based items capturing workplace behaviors. Responses are measured on a five-point scale ranging from ``never'' to ``always.'' Bars represent the percentage of respondents in each category. The sample includes respondents with survey completion times between 194 and 2350 seconds.
\end{minipage}
\end{figure}

Figure~\ref{fig:vignette_freq_combined} reports the distribution of responses to the frequency-based items capturing behavioral dimensions of quiet quitting, measured on five-point frequency scales.

The data indicate that high-frequency disengagement behaviors are rare. For actions involving clear reductions in effort—such as arriving late, calling in sick, or pretending to work—responses are concentrated in the lowest categories. At the same time, other behaviors show more dispersion. In particular, helping colleagues and responding to work matters on days off show substantial mass in intermediate and higher frequency categories.

\begin{figure}[htbp]
\centering
\includegraphics[width=\textwidth]{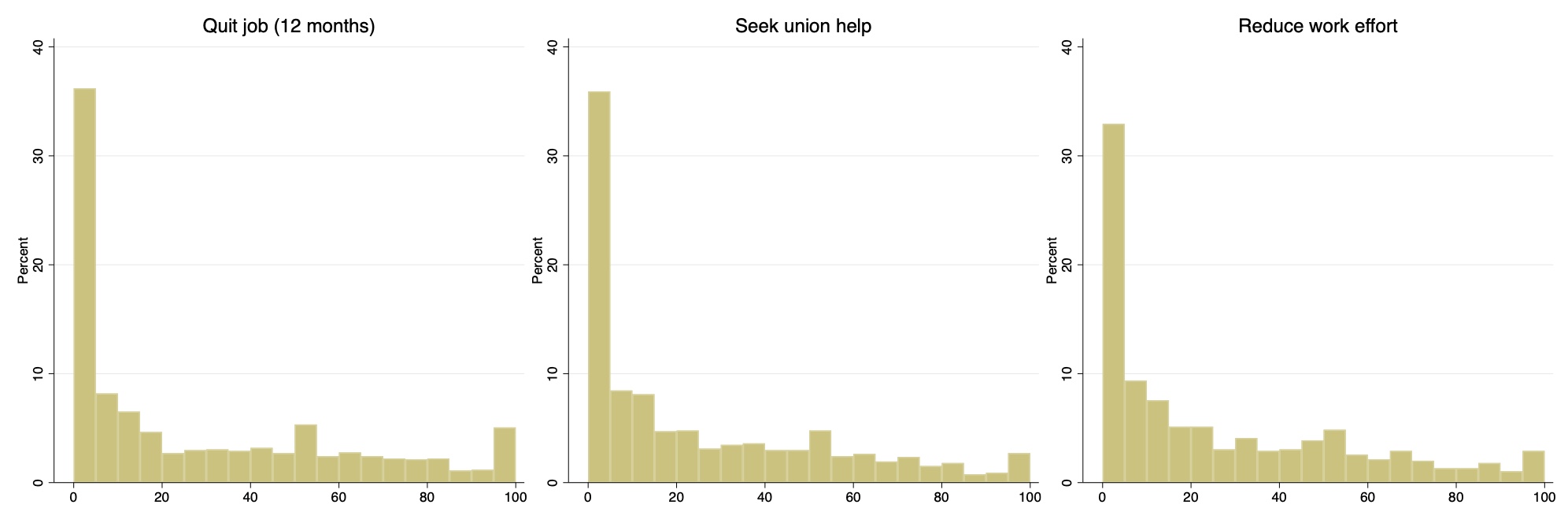}
\caption{Distribution of probabilities of workplace actions}
\label{fig:probability_combined_v3}
\begin{minipage}{\textwidth}
\footnotesize
\textit{Notes:} The figure reports the distribution of subjective probabilities associated with three workplace-related actions: quitting the job within the next 12 months, seeking help from trade unions, and reducing work effort. Responses are expressed on a scale from 0 to 100. Bars represent the percentage of respondents in each bin. The sample includes respondents with survey completion times between 194 and 2350 seconds.
\end{minipage}

\end{figure}

Figure~\ref{fig:probability_combined_v3} reports the distribution of subjective probabilities associated with three workplace-related actions: quitting the job within the next 12 months, seeking help from trade unions, and reducing work effort.

Across all three outcomes, a large mass of respondents assigns a zero probability to the corresponding action. At the same time, a non-negligible share reports strictly positive probabilities, with responses spread over a wide range of values. The distributions also show clustering at focal values such as 50 and 100, indicating substantial heterogeneity in the beliefs reported regarding these actions.

\begin{figure}[htbp]
\centering
\includegraphics[width=0.65\textwidth]{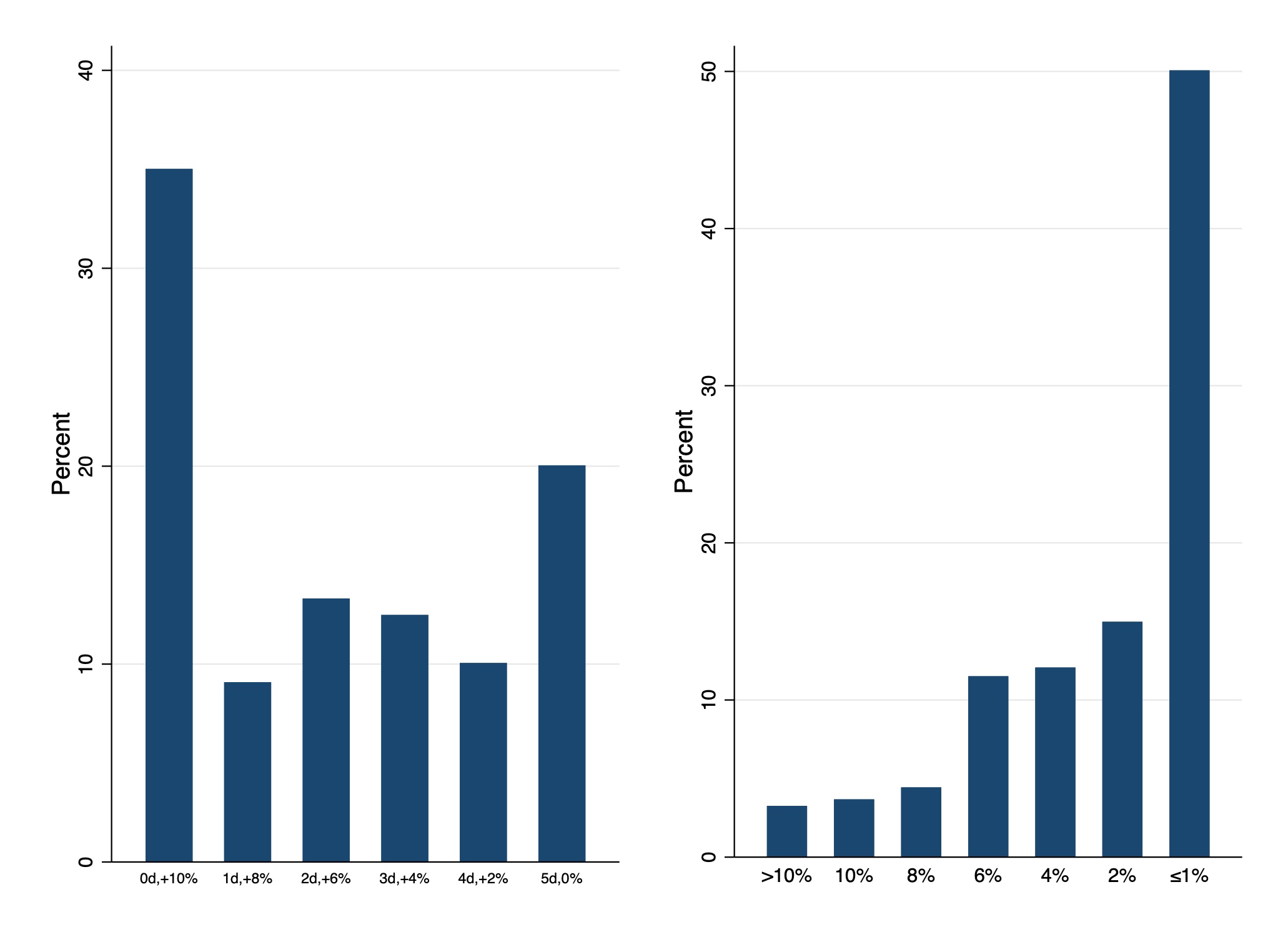}
\caption{Distribution of willingness to accept wage reductions for remote work}
\label{fig:law_combined_v3}
\begin{minipage}{\textwidth}
\footnotesize
\textit{Notes:} The figure reports, on the left, the distribution of respondents’ choices in a scenario where employers provide monetary compensation for work not performed from home, with each option combining remote working days and wage increases; and, on the right, the distribution of respondents’ willingness to give up part of their salary to work one additional day from home, expressed as percentage reductions in net salary. Bars represent the percentage of respondents selecting each option. The sample includes respondents with survey completion times between 194 and 2350 seconds.
\end{minipage}
\end{figure}

Figure~\ref{fig:law_combined_v3} shows, on the left, the distribution of choices in a scenario where a policy requires employers to compensate workers for time not worked from home, with each option combining remote working days and wage increases; and, on the right, the distribution of respondents’ willingness to accept wage reductions in exchange for one additional day of remote work. The most frequently selected option is full on-site work with the highest wage increase, although a substantial share of respondents selects bundles involving remote work, even with lower or no wage premia. Consistently, while many respondents are only willing to give up a small fraction of their salary for additional remote work, a non-negligible share is willing to accept larger reductions, indicating heterogeneity in the valuation of this job amenity.

\begin{figure}[H]
\centering
\includegraphics[width=0.95\textwidth]{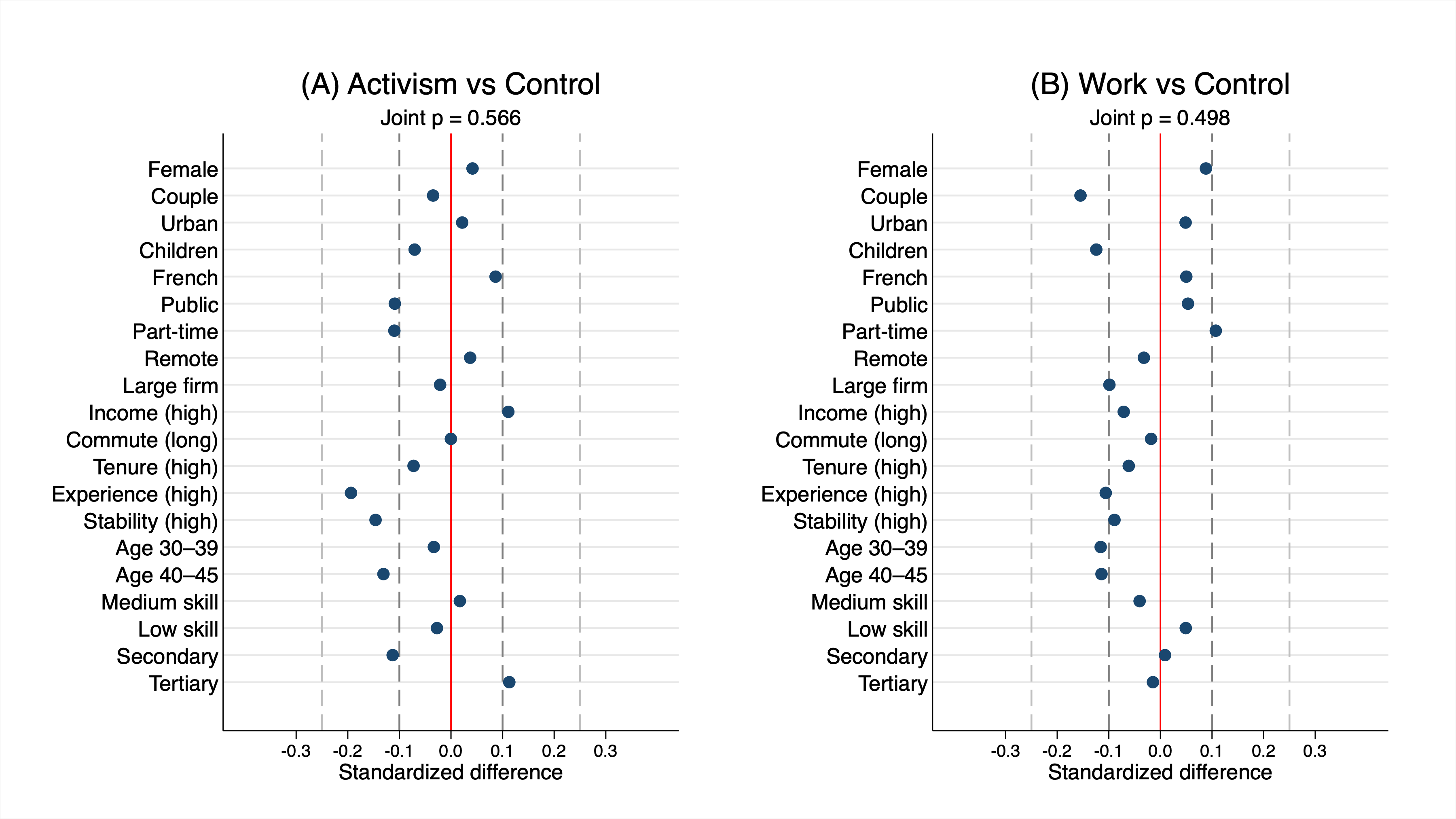}
\caption{Balance across treatment groups}
\label{fig:balance_plot_combined}

\begin{minipage}{0.95\textwidth}
\footnotesize
\begin{spacing}{0.9}
\raggedright
\textit{Notes:} The figure reports standardized differences in baseline characteristics between treatment and control groups. Each point corresponds to the difference in means, scaled by the pooled standard deviation. Vertical lines indicate zero difference (solid) and conventional thresholds of 0.1 and 0.25 in absolute value (dashed). Joint p-values are from tests of the null hypothesis that all covariates are jointly equal across groups. The sample is restricted to respondents with survey completion times between 194 and 2350 seconds.
\end{spacing}
\end{minipage}
\end{figure}

\textbf{Balancing Test}: Figure~\ref{fig:balance_plot_combined} reports standardized differences in baseline characteristics between the treatment and control groups. The estimates are small in magnitude and centered around zero, with no discernible pattern across covariates: all absolutestandardized differences lie below 0.10, the conventional threshold for imbalance.
Consistent with this evidence, joint orthogonality tests, based on a regression of
treatment status on the full set of baseline covariates, fail to reject the null that
all coefficients are jointly zero ($F = \langle \cdot \rangle$, $p = 0.566$ for Activism;
$F = \langle \cdot \rangle$, $p = 0.498$ for Work), consistent with successful
randomization.


\section{Empirical strategy}
\label{sec:empirical}


Our main empirical analysis focuses on comparisons between each treatment and the control condition. The baseline specification estimates average treatment effects (ATEs) by regressing each outcome on treatment indicators and a vector of pre-treatment individual and job-related characteristics. Formally, for outcome $y_i$, we estimate:

\begin{equation}
y_i = \alpha + \beta_1 \mathbbm{1}\{\text{Activism}_i\}
+ \beta_2 \mathbbm{1}\{\text{Work}_i\}
+ \beta_3 \mathbbm{1}\{\text{Work+Activism}_i\}
+ X_i'\gamma + \varepsilon_i,
\end{equation}

where the control condition is the omitted category. The vector $X_i$ includes socio-demographic characteristics, job attributes, and workplace features measured prior to treatment exposure. All standard errors are robust to heteroskedasticity.

Because outcomes differ in their statistical support, we adopt estimation methods suited to their measurement scale. Binary outcomes are estimated using probit models and reported as average marginal effects. Outcomes expressed as probabilities on the unit interval are estimated using fractional probit models, while discrete preference outcomes related to remote work and compensation are estimated using Poisson pseudo-maximum likelihood (PPML). These approaches ensure that predicted values remain within the admissible support of each outcome.

To mitigate concerns related to inattentive responding, we trim the estimation sample by excluding observations with implausibly short or long completion times.

\subsection{Specification with all treatment arms}

Our preferred specification compares each pure treatment to the control condition and excludes the hybrid arm, so that the contrast of interest is not affected by the composition of the residual sample. As a specification check, we also estimate the full version of equation~(1) on the entire sample, retaining the three treatment indicators simultaneously---Activism, Work \& Employment, and the hybrid Work + Activism condition---with the control condition as the omitted category. This specification increases the estimation sample by roughly one third and provides the direct test of H3.

\subsection{Inference with multiple outcomes}

Because within-job disengagement is measured through several items belonging to the same construct, testing each outcome in isolation raises the probability of at least one spurious rejection. We therefore complement conventional $p$-values with family-wise error rate corrections based on the Romano--Wolf stepdown procedure, implemented through a bootstrap with 5{,}000 replications. We define two families: the six individual items of the Quiet Quitting Scale, and the three aggregate indices obtained by averaging the two items loading on each latent factor. Because the two families differ in size, adjusted $p$-values are comparable within but not across families.

\subsection{Heterogeneous treatment effects}

In addition to baseline effects, we explore whether treatment responses vary across predetermined individual and job-related characteristics. All moderators are measured before treatment exposure and are therefore orthogonal to treatment assignment.

We first examine heterogeneity along binary moderators capturing key dimensions of workers’ demographic characteristics, labor market position, and working conditions. These include gender, family status, education, job tenure, labor market experience, firm size, sector of activity, commuting time, and the possibility to work remotely. Continuous variables are discretized using median splits.

For each moderator, we estimate models fully interacted with treatment indicators and compute subgroup-specific average marginal effects. To assess whether treatment effects differ across groups, we report tests of equality between subgroup-specific marginal effects.

We also explore heterogeneity along selected ordinal dimensions using categorical moderators, including age, individual income, and household income. For these specifications, we estimate models fully interacted with treatment indicators and compute marginal effects for each category relative to the control group. Differences across categories are assessed using Wald tests.

\section{Main results}
\label{sec:Results}

\subsection{Baseline treatment effects}

For ease of interpretation, we present the baseline treatment effects graphically in Figure~\ref{fig:baseline_coefplots}, while the corresponding regression estimates are reported in Appendix Tables~\ref{tab:main_vignette} and~\ref{tab:main_prob_law}. Panels (a) and (b) display the effects of the Social Justice \& Activism treatment, whereas panels (c) and (d) report the corresponding estimates for the Work \& Employment treatment. In all specifications, we compare each treatment to the control condition and exclude the hybrid treatment. Regressions include a rich set of socio-demographic and job-related controls, and the estimation sample is trimmed to exclude respondents with implausibly short or long survey completion times.

\begin{figure}[htbp]
    \centering
    \begin{minipage}{0.48\textwidth}
        \centering
        \includegraphics[width=\linewidth]{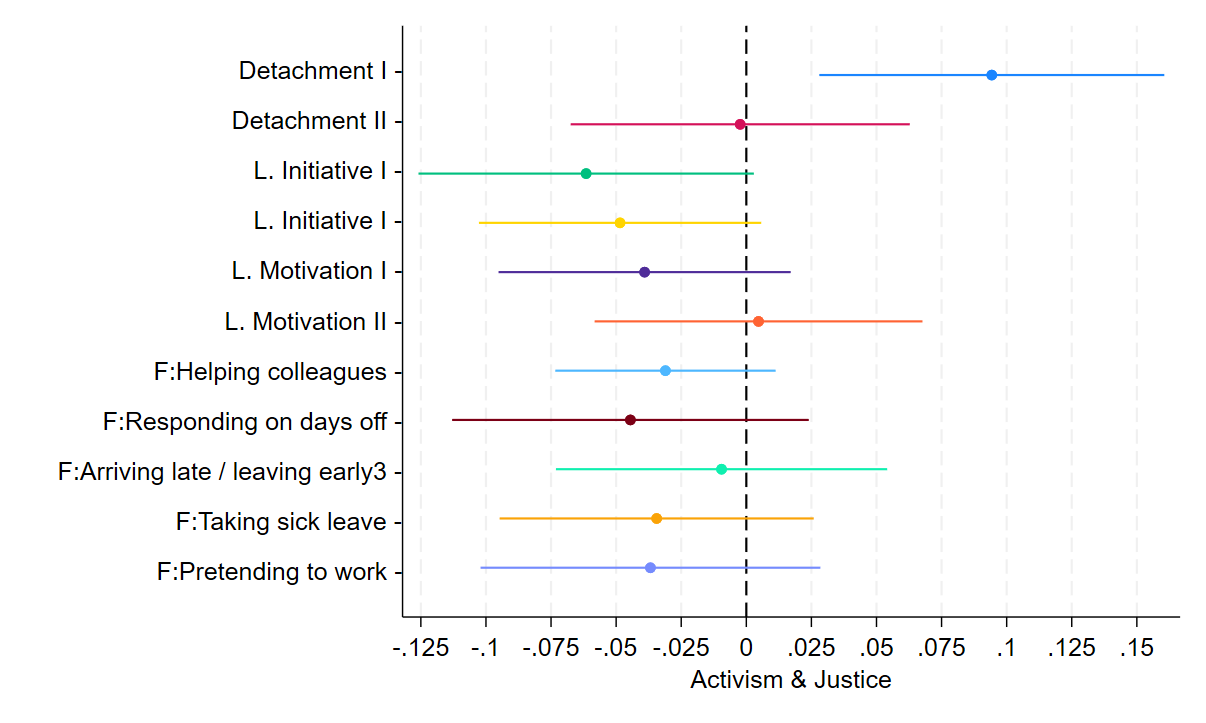}
        \smallskip
        \textit{(a) Activism: Vignette-based and frequency outcomes}
    \end{minipage}
    \hfill
    \begin{minipage}{0.48\textwidth}
        \centering
        \includegraphics[width=\linewidth]{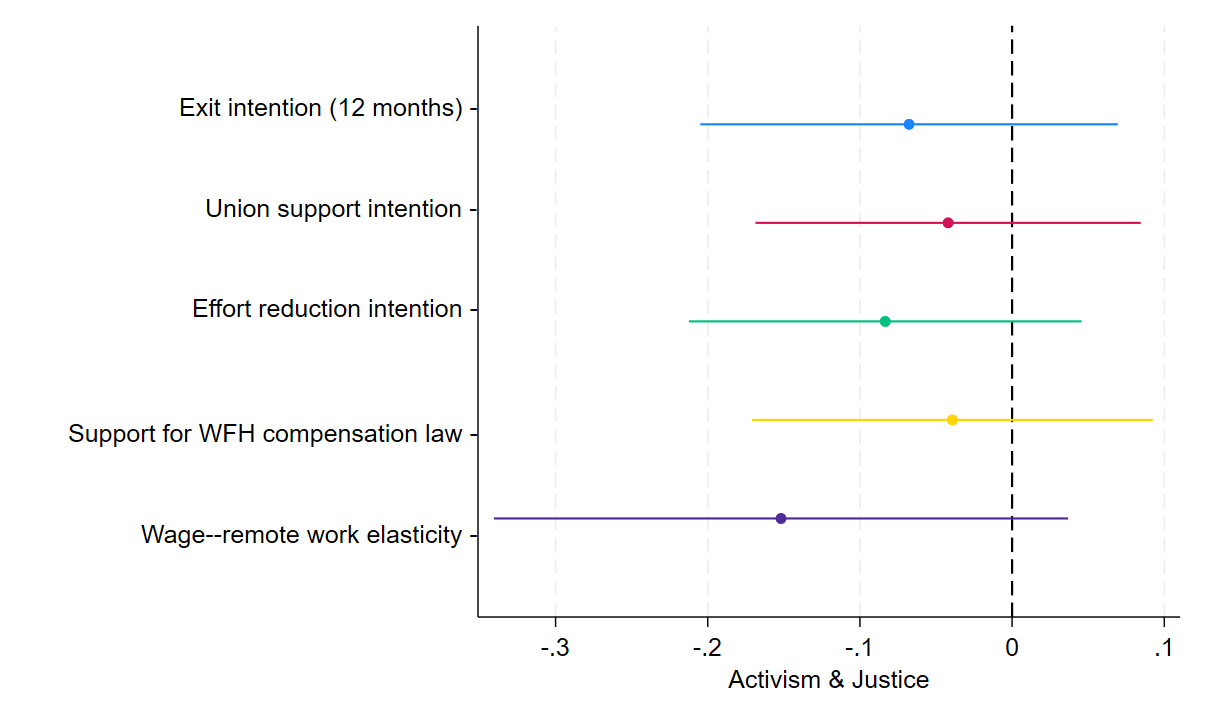}
        \smallskip
        \textit{(b) Activism: Probability, law, and elasticity outcomes}
    \end{minipage}

    \vspace{0.8em}

    \begin{minipage}{0.48\textwidth}
        \centering
        \includegraphics[width=\linewidth]{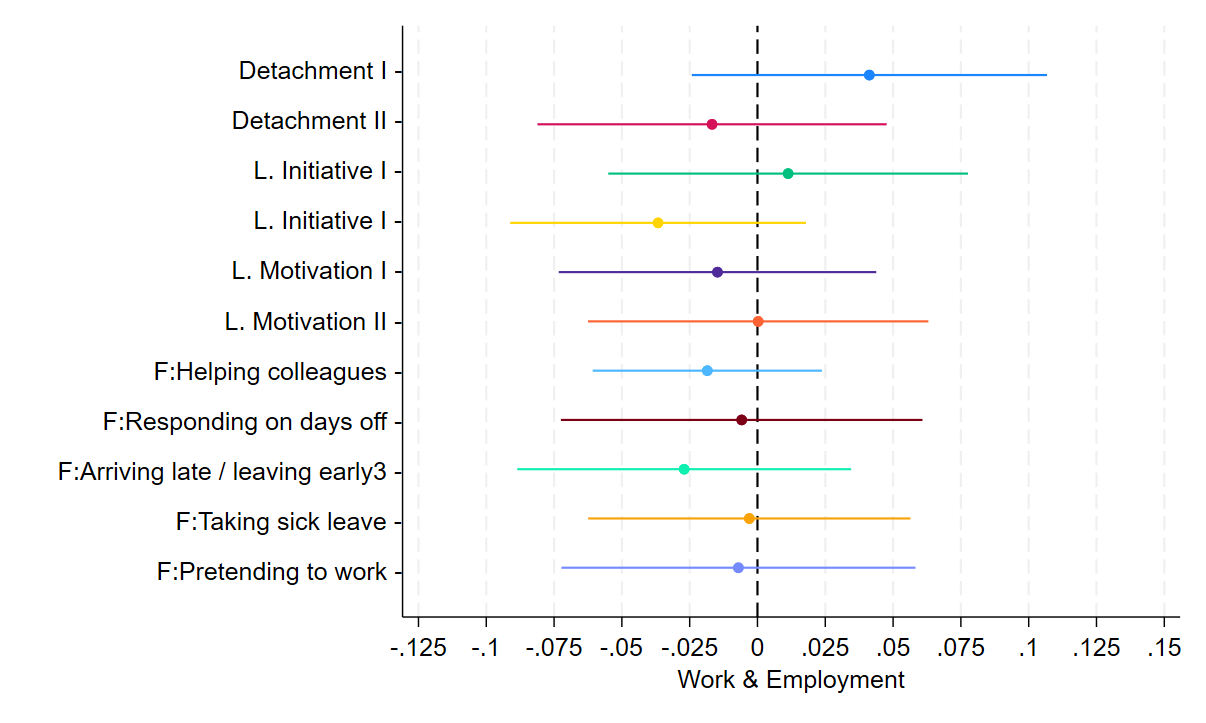}
        \smallskip
        \textit{(c) Work \& Employment: Vignette-based and frequency outcomes}
    \end{minipage}
    \hfill
    \begin{minipage}{0.48\textwidth}
        \centering
        \includegraphics[width=\linewidth]{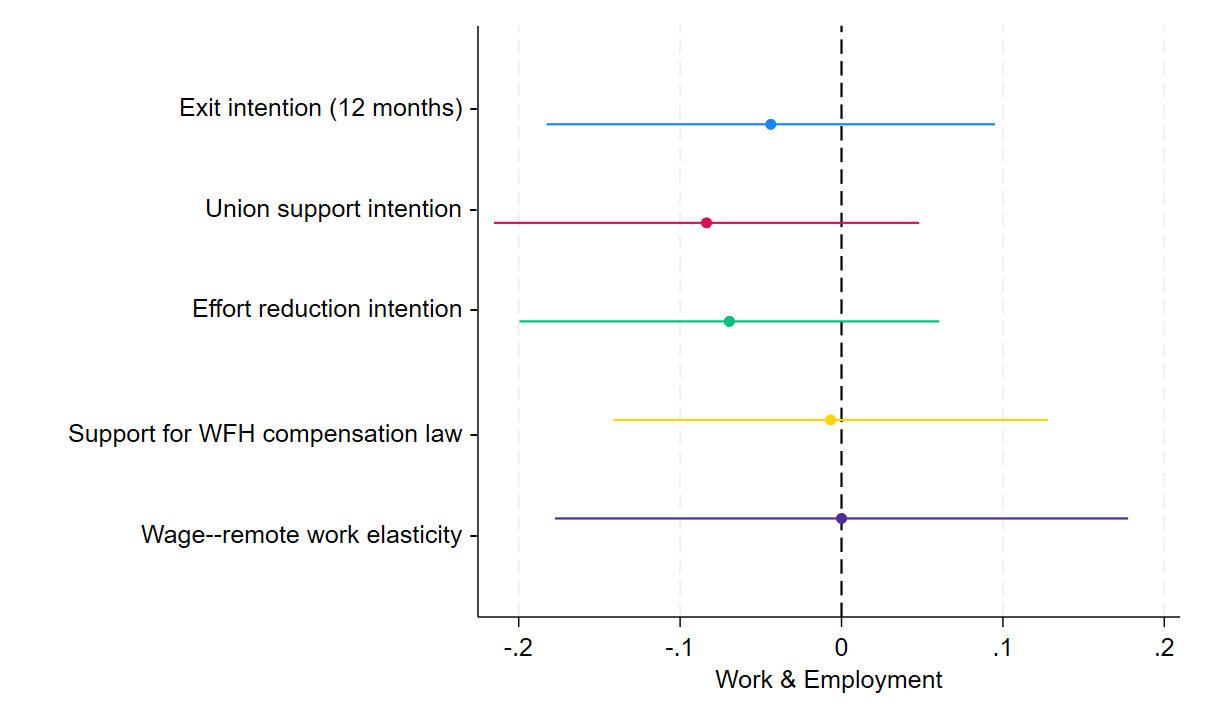}
        \smallskip
        \textit{(d) Work \& Employment: Probability, law, and elasticity outcomes}
    \end{minipage}

    \caption{Baseline treatment effects relative to the control condition}
    \label{fig:baseline_coefplots}

    \vspace{1em}
    \begin{minipage}{0.95\textwidth}
        \scriptsize
        \textit{Notes:} Points represent average marginal effects and bars denote 95\% confidence intervals.All specifications include socio-demographic and job-related controls and exclude the hybrid vignette. The estimation sample is trimmed by survey completion time to mitigate inattentive responding.The control condition (workplace amenities) is the omitted category.
    \end{minipage}
\end{figure}

Panel (a) shows the estimated average marginal effects of the Social Justice \& Activism treatment on agreement with statements capturing detachment, lack of initiative, and lack of motivation, as well as on the reported frequency of disengaged behaviors. A clear and heterogeneous pattern emerges across dimensions of disengagement. In particular, exposure to the treatment increases agreement with items capturing detachment from work by approximately 9 percentage points relative to the control group. At the same time, the treatment significantly reduces agreement with items related to lack of initiative and passive disengagement, with effects ranging between 4 and 6 percentage points in absolute value. These estimates are statistically significant and economically meaningful, suggesting that frames emphasizing social justice and collective action reshape workers’ interpretations of the employment relationship rather than inducing uniform disengagement.

By contrast, panels (c) and (d) show that the Work \& Employment treatment does not generate systematic effects across the same set of outcomes. Estimated coefficients are generally small in magnitude and centered around zero, and none can be statistically distinguished from zero. As reported in Appendix Table~\ref{tab:main_vignette}, this pattern holds both for agreement-based measures and for reported disengaged behaviors, indicating that narratives focused on ethical employment practices and work organization do not substantially alter workers’ stated attitudes or behaviors relative to the control condition.

Panels (b) and (d) turn to broader behavioral intentions, including stated probabilities of quitting, seeking union assistance, and reducing work effort, as well as preferences for regulatory intervention and willingness to trade off wages for additional remote work. Across these outcomes, confidence intervals consistently overlap zero for both treatments. Point estimates for the Social Justice \& Activism treatment are generally negative—on the order of 4 to 8 percentage points for stated behavioral probabilities—but remain imprecisely estimated and cannot be distinguished from zero (Appendix Table~\ref{tab:main_prob_law}). Similarly, we find no significant effects on preferences for regulatory intervention or on the compensation–remote work trade-off.

Taken together, these results indicate that exposure to alternative workplace frames primarily affects attitudinal dimensions of work disengagement without translating into immediate changes in stated quitting behavior, effort reduction, or policy preferences. This pattern is consistent with the view that shifts in norms and perceptions may precede more consequential behavioral responses.

\subsection{Robustness}

\textbf{The hybrid treatment.} Appendix Tables~\ref{tab:main_vignette_3t} and~\ref{tab:main_prob_law_3t} replicate the analysis on the full sample, retaining all three treatment indicators simultaneously. The estimates for the Social Justice \& Activism treatment are essentially unchanged: agreement with the detachment item increases by 9.2 percentage points, while agreement with the items capturing lack of initiative and silence declines by 6.2 and 4.8 percentage points respectively. The hybrid Work + Activism condition reproduces the overt component of this pattern in attenuated form, raising agreement with the detachment item by 7.0 percentage points, but it does not reduce agreement with the passive-disengagement items, whose coefficients are small and positive. The Work \& Employment treatment remains indistinguishable from the control condition throughout. Consistent with H3, combining the two frames does not amplify the effect of the justice-oriented frame: it dilutes it, and in particular it fails to reproduce the reallocation from passive to overt disengagement that characterizes the pure Activism condition. Estimates for behavioral intentions and policy preferences (Appendix Table~\ref{tab:main_prob_law_3t}) remain statistically indistinguishable from zero across all three arms, with the exception of the wage--remote work elasticity under Activism, which is marginally significant in this specification and should be read with caution given the number of outcomes considered.

\textbf{Multiple hypothesis testing.} Because the effects discussed above are estimated over a set of related items, Appendix Table~\ref{tab:mht} reports unadjusted and Romano-Wolf adjusted $p$-values for the Activism treatment. Within the family of six individual scale items, the effect on the detachment item survives the correction (adjusted $p = 0.048$), confirming that the increase in overt disengagement is not an artifact of multiple testing. The two lack-of-initiative items, individually significant at the ten percent level before adjustment, do not survive it (adjusted $p = 0.349$ for both), which is unsurprising given that they load on the same latent factor and therefore split a common signal across two hypotheses. Consistent with this reading, when the items are aggregated into the three indices of the second family, the lack-of-initiative index is significant before adjustment ($p = 0.024$) and remains marginally so afterwards (adjusted $p = 0.067$). The reallocation of disengagement across channels is thus robust on its overt side and suggestive, though not conclusive at conventional thresholds, on its passive side. No effect in the remaining outcome families is significant prior to adjustment, so corrections could only raise those $p$-values further.

In the next section, we examine whether these effects are heterogeneous across workers with different levels of employment stability, bargaining power, and exposure to flexible work arrangements.

\subsection{Heterogeneity by predetermined characteristics}
\begin{figure}[h]
\centering

\begin{minipage}{0.48\textwidth}
    \centering
    \includegraphics[width=\linewidth]{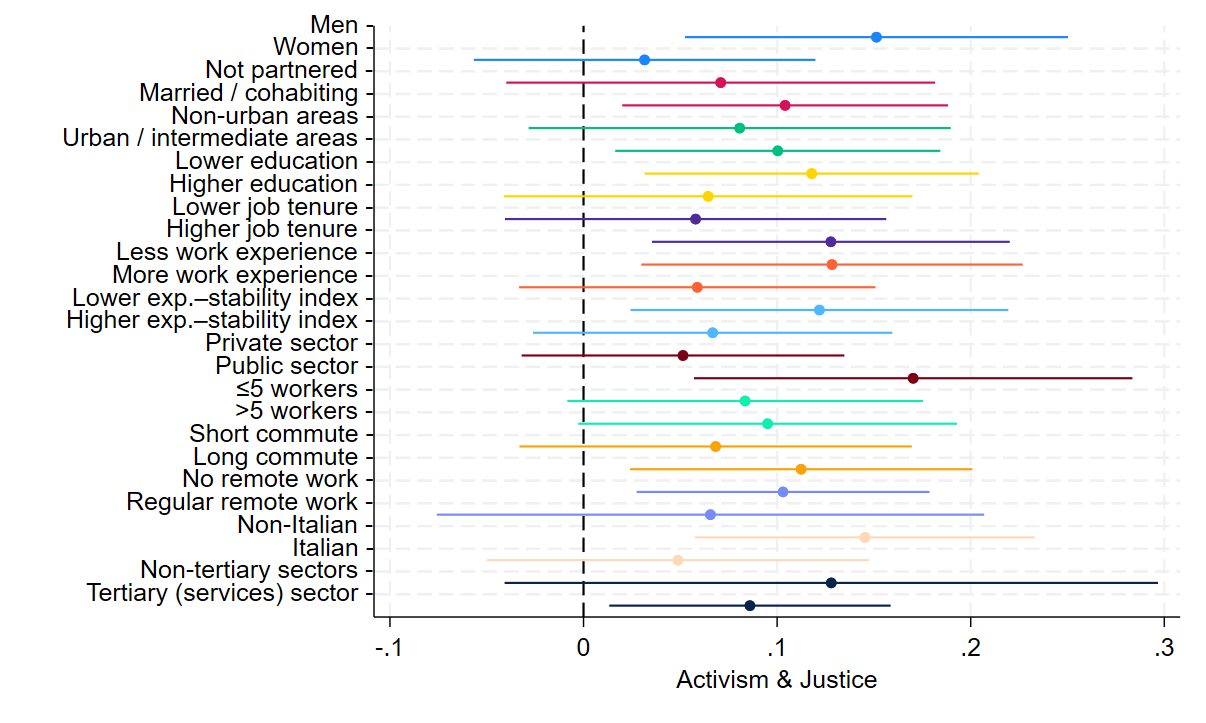}
    
    \vspace{0.3em}
    {\footnotesize (a) Detachment}
\end{minipage}
\hfill
\begin{minipage}{0.48\textwidth}
    \centering
    \includegraphics[width=\linewidth]{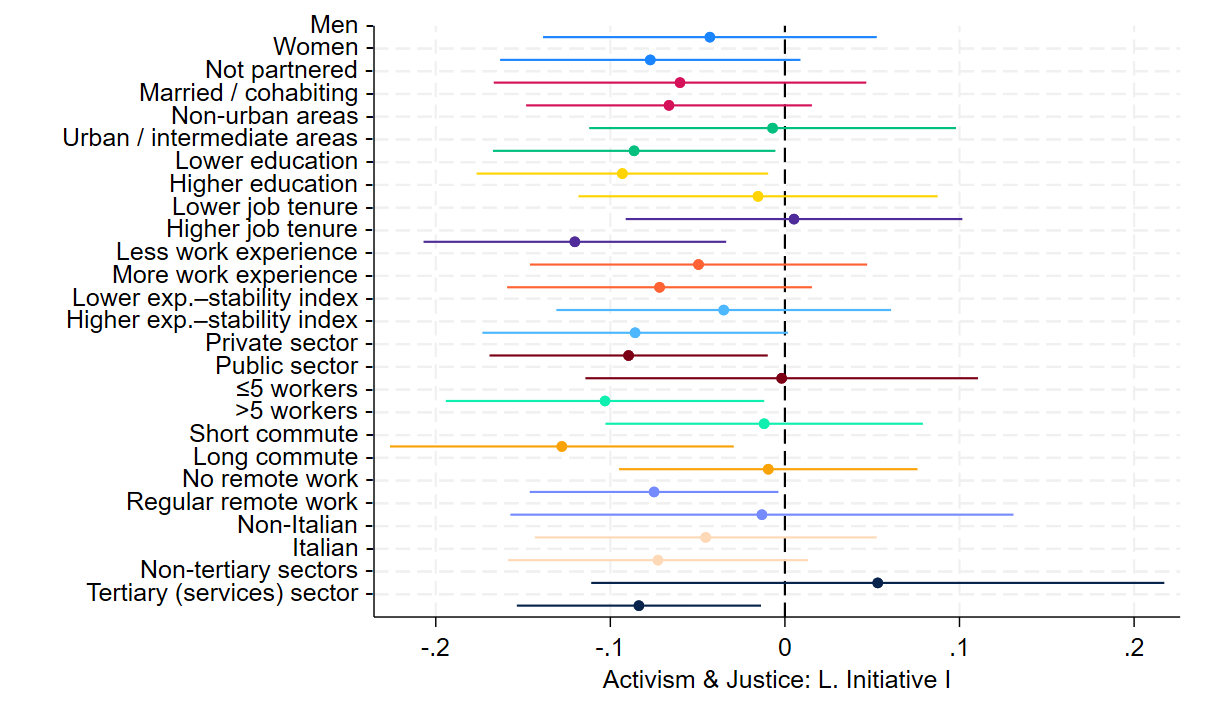}
    
    \vspace{0.3em}
    {\footnotesize (b) Lack of Initiative I}
\end{minipage}

\vspace{0.8em}

\begin{minipage}{0.48\textwidth}
    \centering
    \includegraphics[width=\linewidth]{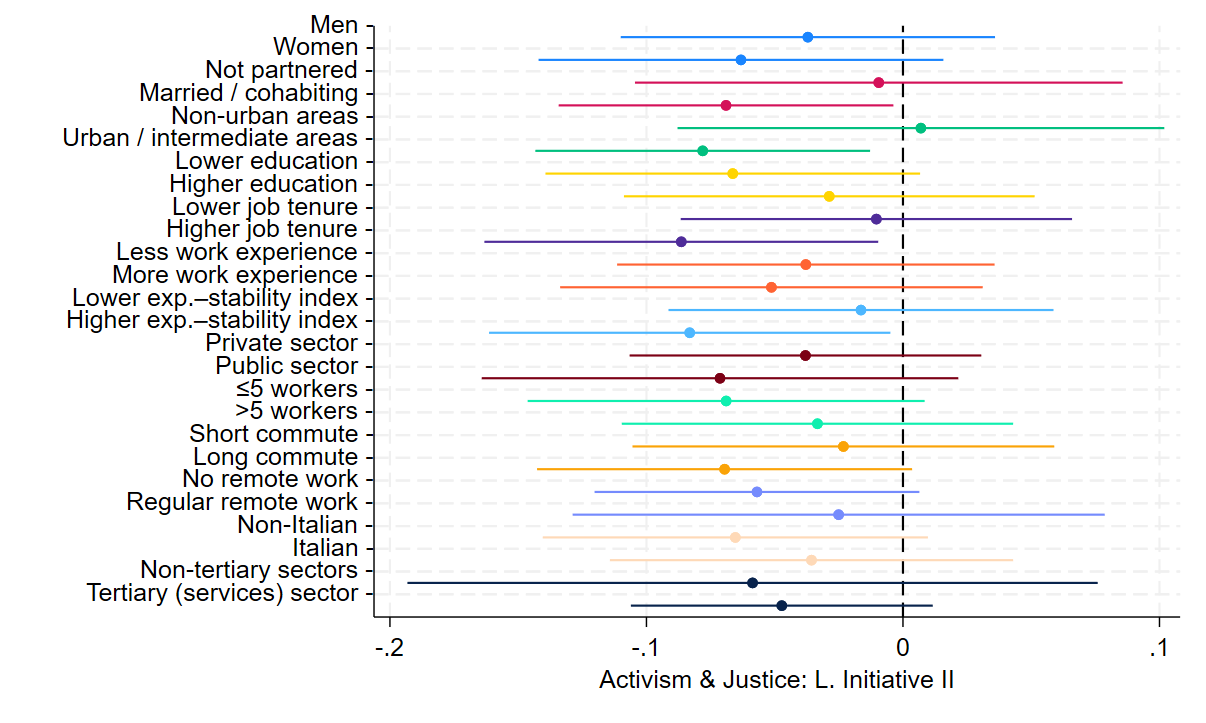}
    
    \vspace{0.3em}
    {\footnotesize (c) Lack of Initiative II}
\end{minipage}
\hfill
\begin{minipage}{0.48\textwidth}
\end{minipage}

\caption{Heterogeneous treatment effects of Activism by predetermined characteristics}
\label{fig:heterogeneity_activism}

\vspace{1em}
\begin{minipage}{0.98\textwidth}
\scriptsize
\textit{Notes.} Each panel reports average marginal effects of the Activism treatment estimated from fully interacted probit models, separately for subgroups defined by predetermined characteristics. Points represent average marginal effects, and horizontal bars indicate 95\% confidence intervals. The vertical dashed line denotes zero effect. All moderators are measured prior to treatment assignment. Each specification includes the full set of baseline controls, excluding the moderator under consideration, as well as interactions between language and geographic and sectoral indicators. Standard errors are robust.
\end{minipage}

\end{figure}

\textbf{Binary Moderators} Figure \ref{fig:heterogeneity_activism} displays heterogeneous treatment effects of the Activism treatment across a set of predetermined \emph{binary} individual and job-related characteristics, while full estimates are reported in Appendix Tables~\ref{tab:heterogeneity_activism} and~\ref{tab:heterogeneity_activism_age_income}. These moderators are constructed using natural dichotomies or median splits and are measured prior to treatment assignment. Overall, heterogeneity patterns are driven primarily by workers’ position within the employment relationship and by contextual constraints, rather than by purely demographic characteristics.


Overt detachment (Detachment I) increases substantially among workers with stronger job protection. For instance, the estimated average marginal effect exceeds 12 percentage points among workers with high job tenure, compared to values close to zero among those with shorter tenure. Similarly, among public sector employees the effect is large and statistically significant (approximately +17 percentage points), while it is small and insignificant in the private sector. Differences by gender are also visible: the treatment increases openly endorsed effort reduction by about 15 percentage points among men, compared to roughly 3 percentage points among women, with the difference being marginally statistically significant.

By contrast, for outcomes capturing passive forms of disengagement (Lack of Initiative I and II), the Activism treatment tends to reduce support for quiet quitting behaviors, but this effect is concentrated among workers facing tighter structural constraints. For example, among respondents with high job tenure, support for reduced effort declines by more than 10 percentage points in Lack of Initiative I, whereas the effect is close to zero for workers with lower tenure. A similar pattern emerges for commuting time: workers facing longer commutes experience a reduction in support for quiet quitting of around 12 percentage points, while no meaningful effect is detected among those with shorter commutes.

Heterogeneity along demographic dimensions such as education, family status, and language is comparatively limited. Although some differences emerge for specific outcomes, they are generally small in magnitude and not statistically distinguishable across groups. Overall, these results suggest that the Activism treatment primarily operates by shifting workers away from fearful, low-visibility disengagement and toward openly declared effort recalibration, particularly in contexts where overt detachment is institutionally or economically less costly, that is, where job protection lowers the expected cost of visibly working to rule.

\begin{figure}[H]
\centering

\begin{minipage}{0.48\textwidth}
    \centering
    \includegraphics[width=\linewidth]{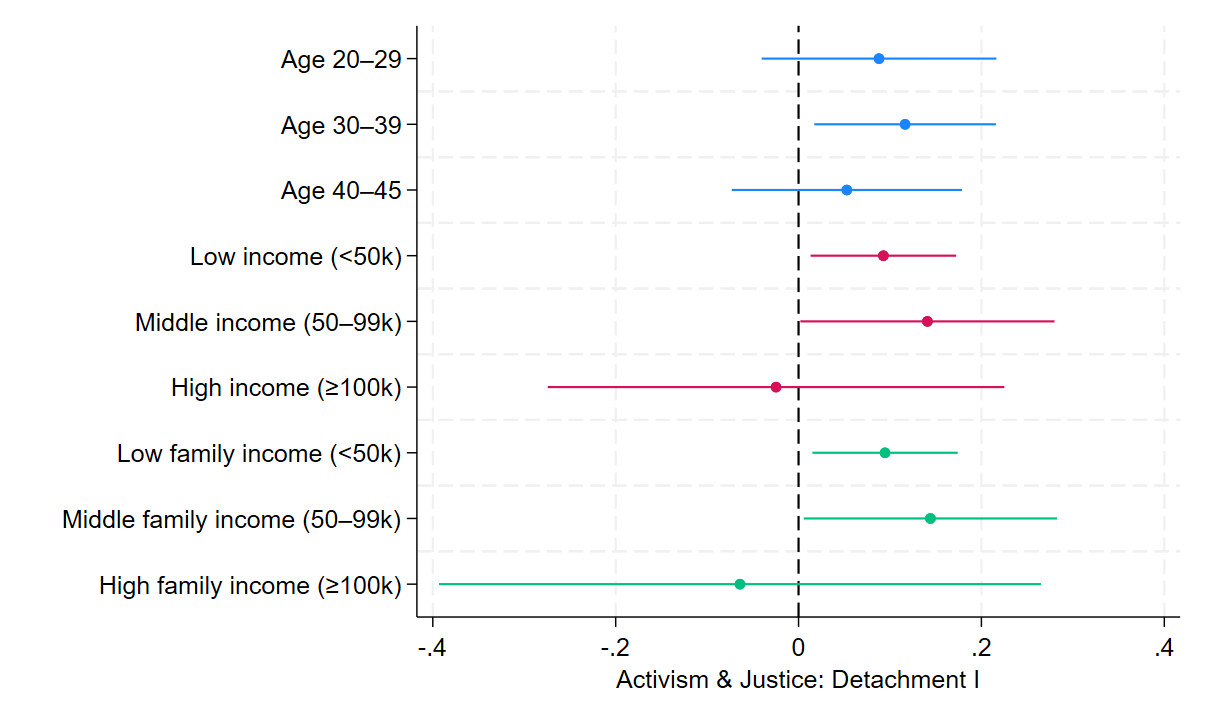}
    
    \vspace{0.3em}
    {\footnotesize (a) Detachment I}
\end{minipage}
\hfill
\begin{minipage}{0.48\textwidth}
    \centering
    \includegraphics[width=\linewidth]{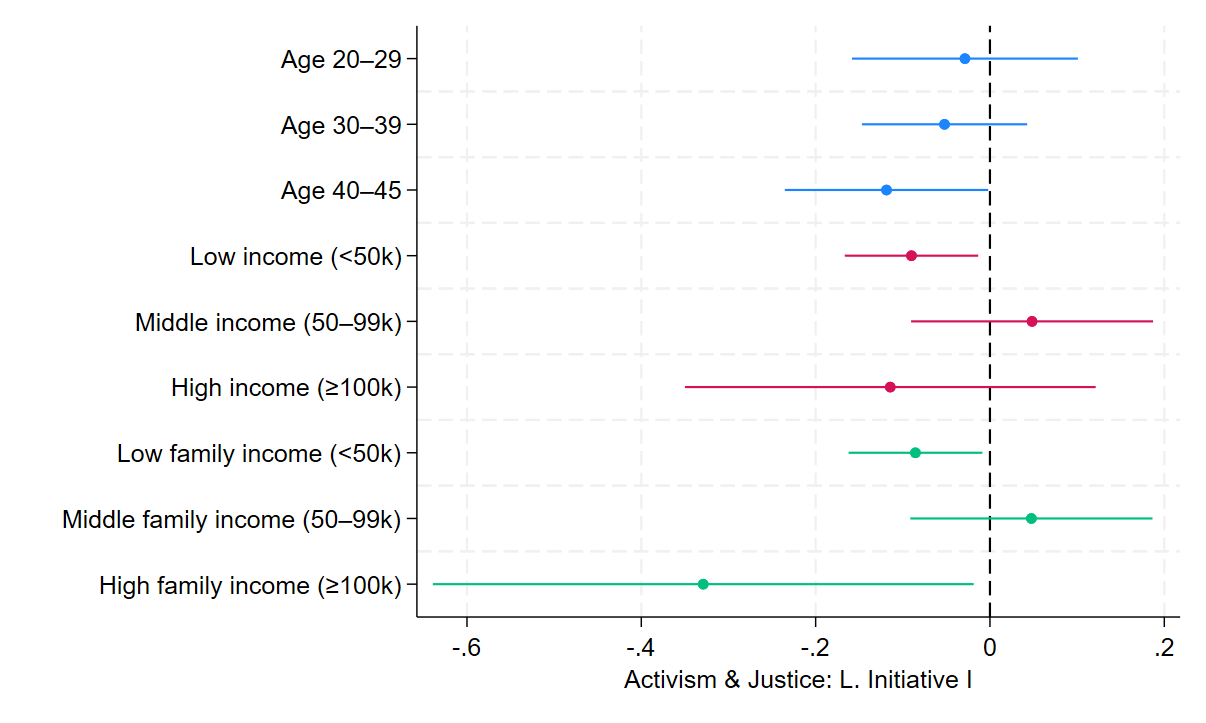}
    
    \vspace{0.3em}
    {\footnotesize (b) Lack of Initiative I}
\end{minipage}

\vspace{0.8em}

\begin{minipage}{0.48\textwidth}
    \centering
    \includegraphics[width=\linewidth]{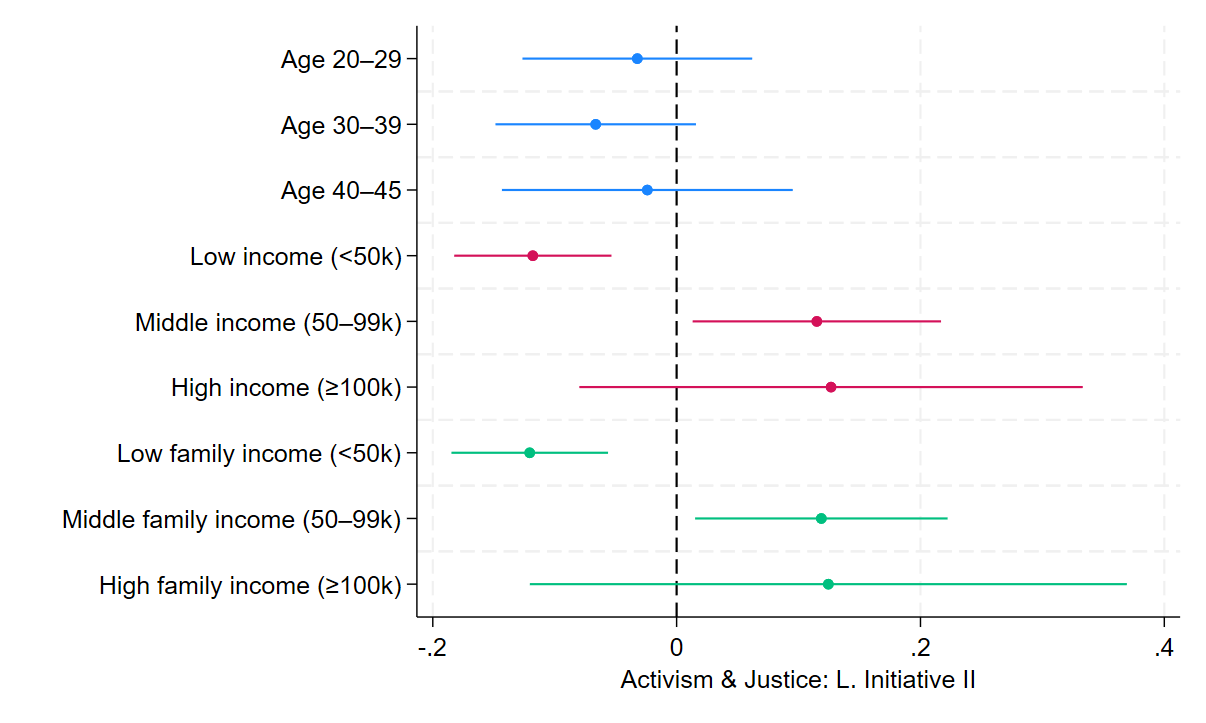}
    
    \vspace{0.3em}
    {\footnotesize (c) Lack of Initiative II}
\end{minipage}
\hfill
\begin{minipage}{0.48\textwidth}
\end{minipage}

\caption{Heterogeneous treatment effects of the Activism treatment by categorical moderators}
\label{fig:heterogeneity_activism_cat}

\vspace{1em}
\begin{minipage}{0.98\textwidth}
\scriptsize
\textit{Notes.} Each panel reports average marginal effects of the Activism treatment estimated from fully interacted probit models with categorical moderators for age, individual income, and household income. Points represent average marginal effects for each category, and horizontal bars indicate 95\% confidence intervals. The vertical dashed line denotes zero effect. All moderators are measured prior to treatment assignment. Each specification includes the full set of baseline controls, excluding the categorical moderator under consideration, as well as interactions between language and geographic and sectoral indicators. Standard errors are robust.
\end{minipage}

\end{figure}

\textbf{Categorical moderators.} We further explore heterogeneity using categorical definitions of age, individual income, and household income. Figure~\ref{fig:heterogeneity_activism_cat} visualizes the corresponding treatment effects, while full estimates and omnibus tests are reported in Appendix Table~\ref{tab:heterogeneity_activism_age_income}. These specifications allow us to assess whether treatment effects vary monotonically or exhibit nonlinear patterns across ordered groups, beyond the binary splits considered above.

Across age groups, treatment effects are relatively stable and omnibus tests consistently fail to reject equality across categories for all outcomes. While point estimates vary somewhat, for example, overt detachment in Detachment I is highest among middle-aged respondents (approximately +12 percentage points), there is no evidence of sharp age-related discontinuities in treatment responses. By contrast, income-related heterogeneity is more pronounced, though not for the outcome that drives our main results. For Detachment I, the estimated effect is positive and significant in the bottom two terciles of individual income ($0.093$, $p=0.022$; $0.141$, $p=0.047$) and indistinguishable from zero in the top tercile ($-0.025$, $p=0.847$), yet the omnibus test cannot reject equality across groups ($p=0.519$). We therefore do not read this ordering as evidence of an income gradient in overt disengagement, if anything, it runs counter to an interpretation of income as a proxy for the job security that drives the tenure and sector results. Significant heterogeneity emerges instead for Lack of Initiative II, where the effect is negative in the bottom tercile ($-0.118$, $p<0.01$) and positive in the middle tercile ($0.115$, $p=0.027$), and the omnibus test rejects equality ($p<0.01$). Estimates based on household income are almost identical throughout, which is unsurprising given the correlation between the two measures and
should not be read as independent confirmation.


Taken together, these results suggest that income-related constraints play an important role in shaping responses to the Activism treatment, particularly for outcomes capturing passive disengagement. At the same time, the absence of strong age-related heterogeneity indicates that the patterns documented using binary moderators are not driven by life-cycle effects but rather by structural economic conditions associated with income and resources.

\section{Conclusions}
\label{sec:Conclusions}

This paper studies how workers respond to workplace dissatisfaction when exit is not the primary margin of adjustment. Using a randomized survey experiment, we show that alternative moral framings of the employment relationship do not uniformly increase or decrease disengagement, but instead reallocate workers’ responses across behavioral margins.

Our main finding is that justice-oriented narratives change the form that disengagement takes rather than its direction: exposure to the Activism treatment increases openly endorsed effort reduction while decreasing fear-based silence and lack of initiative, particularly among workers in more protected employment relationships. Disengagement, in other words, becomes overt rather than hidden. We find no corresponding shift toward collective voice: stated intentions to seek union support are unaffected.

The composition of disengagement, and not only its level, carries distinct implications
for the parties involved. For employers, passive withdrawal is costly precisely because it is unobservable: withheld initiative and silence generate no signal on which management can act, so the loss is both productive and informational. Overt disengagement, by contrast, is detectable, and can in principle be priced and bargained over; the frame converts a hidden information problem into an open contracting one, with the symmetric risk that openly declared effort reduction is more easily normalized among peers. For workers, silence sustained by fear of retaliation is self-reinforcing, since the conditions that motivate it are never disclosed, whereas openly recalibrated effort alters the bargaining posture within the match without incurring the costs of exit, a margin that our results show remains unaffected in the short run. For unions and policymakers, the pattern describes an individualized rather than a collectivized form of conflict: justice-oriented narratives legitimize open individual withdrawal without channelling it into representation, which is informative both for organizing efforts, since latent disaffection becomes observable, and for interventions that presuppose collective voice as the operative margin. Finally, the distinction matters for
measurement. Self-reported quiet quitting captures in part workers' willingness to declare disengagement, not only its underlying level, so a shift in what is considered legitimate to admit can raise measured quiet quitting with limited change in effort actually supplied. This qualification applies to our own stated outcomes as well, and is consistent with our interpretation: what the Activism treatment reallocates is the form in which disaffection is expressed. It also suggests caution in reading post-pandemic trends in survey-based engagement measures as unambiguous evidence of declining effort.


Taken together, these results suggest that workplace dissatisfaction is not a unidimensional phenomenon, but rather a reallocation problem across effort, voice, and exit margins. Moral interpretations of the employment relationship play a central role in shaping which of these margins is activated. In this sense, quiet quitting should not be understood as a fixed behavioral response, but as one of several possible adjustments that depend on how workers interpret their situation.

More broadly, our findings speak to a growing literature emphasizing the role of non-wage job attributes, fairness, and social norms in shaping labor supply decisions. While much of the existing evidence focuses on how these factors affect job choice and mobility, our results highlight their importance within ongoing employment relationships. Changes in narratives and perceived norms can reshape behavior even in the absence of changes in wages or contractual conditions.

Our results also have implications for the interpretation of recent labor market dynamics. While public discourse has emphasized rising quits during the Great Resignation, our findings suggest that a substantial share of adjustment may instead occur within jobs, through shifts in effort provision and forms of worker voice.

Finally, the absence of short-run effects on quitting intentions suggests that changes in norms and interpretations may precede more consequential labor market decisions. Understanding how such shifts translate into longer-run behavior remains an important avenue for future research. In particular, linking moral framings to actual workplace behavior and job mobility using administrative or field data would provide a promising direction to further assess the economic relevance of these mechanisms.

Overall, this paper highlights the importance of moral and interpretive dimensions of work in shaping labor supply along the intensive margin. By showing that narratives can reallocate behavior across effort, voice, and exit, our findings contribute to a more nuanced understanding of worker responses in contemporary labor markets.
\newpage

\bibliographystyle{aer}
\bibliography{bibliography}

\clearpage
\appendix
\singlespacing

\clearpage

\begin{center}
	{\Large \textbf{Appendix}}
\end{center}

\vspace{1em}

\setcounter{section}{0}
\setcounter{table}{0}
\setcounter{figure}{0}
\setcounter{equation}{0}

\renewcommand{\thesection}{\Alph{section}}
\renewcommand{\thetable}{A\arabic{table}}
\renewcommand{\thefigure}{B\arabic{figure}}
\renewcommand{\theequation}{A\arabic{equation}}

\section{Tables}
\label{app:tables}


\begin{table}[htbp]\centering
	\caption{Manipulation check: Relative salience of treatment-specific dimensions}
	\label{tab:manipulation_check}
	
	\setlength{\tabcolsep}{14pt}
	\renewcommand{\arraystretch}{1.15}
	
	\begin{tabular}{lcc}
		\hline\hline
		& (1) & (2) \\
		& Baseline & + Controls \\
		\hline
		WorkTreat & -0.697\sym{***} & -0.733\sym{***} \\
		& (0.108)         & (0.121)         \\
		\hline
		Controls     & No    & Yes   \\
		Observations & 717   & 697   \\
		R-squared    & 0.056 & 0.196 \\
		\hline\hline
	\end{tabular}
	
	\vspace{0.4em}
	
	\begin{minipage}{0.82\textwidth}
		\footnotesize
		\textit{Notes:} The dependent variable is the standardized relative salience gap between treatment-specific keywords and common workplace dimensions, constructed from the post-treatment ranking task. Higher values indicate greater prioritization of channel-consistent dimensions relative to shared individual dimensions. The omitted category is the Activism treatment. Column (2) includes the full set of socio-demographic and job-related controls as described in the empirical strategy section. Robust standard errors are reported in parentheses. \sym{***} \(p<0.01\).
	\end{minipage}
\end{table}

\vspace{5em}

\begin{table}[htbp]\centering
\caption{Two-treatments effects on vignette-based and frequency outcomes}
\label{tab:main_vignette}
\resizebox{\textwidth}{!}{%
\begin{tabular}{l*{12}{c}}
\hline\hline
                & (1) & (2) & (3) & (4) & (5) & (6) & (7) & (8) & (9) & (10) & (11) & (12) \\
\hline
ActivismTreat   & -0.001 & -0.039 & 0.005 & 0.094\sym{***} & -0.062\sym{*} & -0.002 & -0.049\sym{*} & -0.031 & -0.045 & -0.010 & -0.034 & -0.037 \\
                & (0.026) & (0.029) & (0.032) & (0.034) & (0.033) & (0.033) & (0.028) & (0.022) & (0.035) & (0.033) & (0.031) & (0.033) \\
WorkTreat       & -0.005 & -0.015 & 0.000 & 0.041 & 0.011 & -0.017 & -0.037 & -0.019 & -0.006 & -0.027 & -0.003 & -0.007 \\
                & (0.025) & (0.030) & (0.032) & (0.033) & (0.034) & (0.033) & (0.028) & (0.022) & (0.034) & (0.031) & (0.030) & (0.033) \\
\hline
Observations    & 1051 & 1034 & 1047 & 1069 & 1064 & 1071 & 1055 & 1045 & 1065 & 1062 & 1059 & 1066 \\
\hline\hline
\end{tabular}
}
\vspace{0.4em}
\begin{minipage}{0.99\textwidth}
\footnotesize
\textit{Notes:} Entries report average marginal effects from probit models. Standard errors are in parentheses. 
All specifications include socio-demographic and job-related controls and exclude the hybrid vignette. 
The estimation sample is trimmed by survey completion time. 
The control condition (workplace amenities) is the omitted category.
\end{minipage}
\end{table}


\begin{table}[htbp]\centering
	\caption{Two-treatments effects on behavioral intentions and policy preferences}
	\label{tab:main_prob_law}
	
	\small
	\setlength{\tabcolsep}{9pt}
	\renewcommand{\arraystretch}{1.10}
	
	\begin{tabular}{lccccc}
		\hline\hline
		& (1) & (2) & (3) & (4) & (5) \\
		& Quit & Union help & Reduce effort & Law & Elasticity \\
		\hline
		ActivismTreat   & -0.068 & -0.042 & -0.083 & -0.039 & -0.152 \\
		& (0.070) & (0.065) & (0.066) & (0.067) & (0.096) \\
		WorkTreat       & -0.044 & -0.084 & -0.069 & -0.007 & 0.000 \\
		& (0.071) & (0.067) & (0.066) & (0.069) & (0.091) \\
		\hline
		Observations    & 1075 & 1075 & 1075 & 1074 & 1069 \\
		\hline\hline
	\end{tabular}
	
	\vspace{0.4em}
	
	\begin{minipage}{0.85\textwidth}
		\footnotesize
		\textit{Notes:} Columns (1)--(3) report fractional probit estimates for outcomes expressed as probabilities on the unit interval. Columns (4) and (5) report Poisson pseudo-maximum likelihood estimates. Standard errors are reported in parentheses. All specifications include socio-demographic and job-related controls and exclude the hybrid vignette. The control condition (workplace amenities) is the omitted category.
	\end{minipage}
\end{table}


\begin{table}[htbp]\centering
	\caption{Three-treatments effects on vignette-based and frequency outcomes}
	\label{tab:main_vignette_3t}
	\resizebox{\textwidth}{!}{%
		\begin{tabular}{l*{12}{c}}
			\hline\hline
			& (1) & (2) & (3) & (4) & (5) & (6) & (7) & (8) & (9) & (10) & (11) & (12) \\
			\hline
			ActivismTreat   &  0.00339         &  -0.0311         &-0.0000243         &   0.0924\sym{***}&  -0.0623\sym{*}  &  0.00327         &  -0.0479\sym{*}  &  -0.0316         &  -0.0472         & -0.00574         &  -0.0258         &  -0.0272         \\
			& (0.0258)         & (0.0287)         & (0.0320)         & (0.0342)         & (0.0327)         & (0.0336)         & (0.0279)         & (0.0224)         & (0.0354)         & (0.0327)         & (0.0310)         & (0.0333)         \\
			WorkActiveTreat & -0.00832         &  0.00966         &  0.00394         &   0.0698\sym{**} &   0.0205         &  0.00106         &   0.0101         &  -0.0233         &   0.0452         &  0.00304         &  0.00327         &  0.00574         \\
			& (0.0264)         & (0.0304)         & (0.0326)         & (0.0343)         & (0.0337)         & (0.0332)         & (0.0298)         & (0.0223)         & (0.0345)         & (0.0327)         & (0.0314)         & (0.0338)         \\
			WorkTreat       &  -0.0116         &  -0.0183         &  0.00859         &   0.0297         &  0.00610         &  -0.0246         &  -0.0322         &  -0.0101         & -0.00653         &  -0.0287         & -0.00345         & -0.00868         \\
			& (0.0259)         & (0.0293)         & (0.0320)         & (0.0335)         & (0.0337)         & (0.0331)         & (0.0283)         & (0.0215)         & (0.0346)         & (0.0315)         & (0.0302)         & (0.0331)         \\
			\hline
			Observations    &     1387         &     1383         &     1399         &     1411         &     1411         &     1411         &     1389         &     1386         &     1405         &     1411         &     1411         &     1411         \\
			\hline\hline
		\end{tabular}
	}
	\vspace{0.4em}
	\begin{minipage}{0.98\textwidth}
		\footnotesize
		\textit{Notes:} Entries report average marginal effects from probit models. Standard errors are in parentheses.
		All specifications include socio-demographic and job-related controls and retain all three treatment arms
		simultaneously, including the hybrid Work + Activism condition (\textit{WorkActiveTreat}).
		The estimation sample is trimmed by survey completion time.
		The control condition (workplace amenities) is the omitted category.
	\end{minipage}
\end{table}


\begin{table}[htbp]\centering
	\caption{Three-treatments effects on behavioral intentions and policy preferences}
	\label{tab:main_prob_law_3t}
	
	\small
	\setlength{\tabcolsep}{9pt}
	\renewcommand{\arraystretch}{1.10}
	
	\begin{tabular}{lccccc}
		\hline\hline
		& (1) & (2) & (3) & (4) & (5) \\
		& Quit & Union help & Reduce effort & Law & Elasticity \\
		\hline
		ActivismTreat   &  -0.0610         &  -0.0305         &  -0.0676         &  -0.0474         &   -0.155\sym{*}  \\
		& (0.0699)         & (0.0653)         & (0.0656)         & (0.0664)         & (0.0939)         \\
		WorkActiveTreat &  -0.0788         &  -0.0970         &  -0.0611         &   0.0583         &   0.0131         \\
		& (0.0716)         & (0.0661)         & (0.0676)         & (0.0659)         & (0.0902)         \\
		WorkTreat       &  -0.0498         &  -0.0827         &  -0.0640         &  -0.0270         & -0.00331         \\
		& (0.0709)         & (0.0672)         & (0.0664)         & (0.0675)         & (0.0888)         \\
		\hline
		Observations    &     1415         &     1415         &     1415         &     1414         &     1411         \\
	\hline\hline
	\end{tabular}
	
	\vspace{0.4em}
	
	\begin{minipage}{0.91\textwidth}
		\footnotesize
		\textit{Notes:} Columns (1)--(3) report fractional probit estimates for outcomes expressed as probabilities on the unit interval. Columns (4) and (5) report Poisson pseudo-maximum likelihood estimates. Standard errors are reported in parentheses. All specifications include socio-demographic and job-related controls and retain all three treatment arms simultaneously, including the hybrid Work + Activism condition (\textit{WorkActiveTreat}). The control condition (workplace amenities) is the omitted category.
	\end{minipage}
\end{table}



\begin{table}[htbp]
	\centering
	\caption{Multiple hypothesis testing adjustments: quiet quitting outcomes}
	\label{tab:mht}
	\small
	\begin{tabular}{l cc}
		\toprule
		Outcome & Unadjusted $p$ & Romano--Wolf $p$ \\
		\midrule
		\multicolumn{3}{l}{\emph{Panel A. Individual items (family size = 6)}} \\
		\addlinespace
		Detachment I (minimum work)     & 0.009 & 0.048 \\
		Detachment II (breaks)          & 0.999 & 1.000 \\
		L.\ Initiative I (silence)      & 0.085 & 0.349 \\
		L.\ Initiative II (initiative)  & 0.096 & 0.349 \\
		L.\ Motivation I                & 0.249 & 0.579 \\
		L.\ Motivation II               & 0.995 & 1.000 \\
		\addlinespace
		\multicolumn{3}{l}{\emph{Panel B. Aggregate indices (family size = 3)}} \\
		\addlinespace
		Detachment                      & 0.118 & 0.211 \\
		Lack of initiative              & 0.024 & 0.067 \\
		Lack of motivation              & 0.477 & 0.477 \\
		\bottomrule
	\end{tabular}
	
	\vspace{0.5em}
	\begin{minipage}{\textwidth}
		\footnotesize
		\emph{Notes:} The table reports unadjusted and family-wise adjusted $p$-values
		for the effect of the Social Justice \& Activism treatment relative to the
		control condition. Point estimates and standard errors are in
		Table~\ref{tab:main_vignette}. Both columns are computed from the same bootstrap procedure with
		5{,}000 replications, so that the only difference between them is the
		multiplicity adjustment: the first column reports resampling-based $p$-values
		for each hypothesis in isolation, the second reports Romano--Wolf stepdown
		$p$-values controlling the family-wise error rate within each family. The
		indices in Panel B are simple averages of the two items loading on each latent
		factor of the Quiet Quitting Scale (Galanis et al., 2023); because the two
		panels imply families of different size, adjusted $p$-values are not directly
		comparable across panels. The benchmark item ``I give my best at work'' is
		excluded, as it is not part of the quiet-quitting construct. Adjustments are
		not reported for the Work \& Employment treatment, nor for the
		frequency-based, intention, and preference outcome families, because no effect
		in those families is significant at conventional levels prior to adjustment;
		corrections can only raise those $p$-values further. All specifications
		include the full set of socio-demographic and job-related controls and exclude
		the hybrid vignette; the estimation sample is trimmed by survey completion
		time.
	\end{minipage}
\end{table}


\begin{table}[H]\centering
	\caption{Heterogeneous treatment effects by predetermined characteristics --- Activism}
	\label{tab:heterogeneity_activism}
	
	{\footnotesize
		\setlength{\tabcolsep}{5pt}
		\renewcommand{\arraystretch}{0.78}
		
		\textit{Panel A. Individual and demographic characteristics}
		
		\vspace{0.2em}
		
		\begin{tabular}{lccc}
			\hline\hline
			& Vignette 4 & Vignette 5 & Vignette 7 \tabularnewline
			\hline
			Female = 0 & .151*** & -.043 & -.037 \tabularnewline
			& (.003) & (.378) & (.319) \tabularnewline
			Female = 1 & .032 & -.077* & -.063 \tabularnewline
			& (.483) & (.079) & (.117) \tabularnewline
			Difference & -.120* & -.034 & -.026 \tabularnewline
			& (.075) & (.601) & (.631) \tabularnewline
			\hline
			In a couple = 0 & .071 & -.060 & -.009 \tabularnewline
			& (.210) & (.270) & (.846) \tabularnewline
			In a couple = 1 & .104** & -.066 & -.069** \tabularnewline
			& (.015) & (.112) & (.038) \tabularnewline
			Difference      & .033 & -.006 & -.060 \tabularnewline
			& (.639) & (.926) & (.309) \tabularnewline
			\hline
			Urban area = 0 & .081 & -.007 & .007 \tabularnewline
			& (.147) & (.895) & (.885) \tabularnewline
			Urban area = 1 & .100** & -.086** & -.078** \tabularnewline
			& (.019) & (.036) & (.019) \tabularnewline
			Difference     & .020 & -.079 & -.085 \tabularnewline
			& (.779) & (.238) & (.145) \tabularnewline
			\hline
			High education = 0 & .118*** & -.093** & -.066* \tabularnewline
			& (.007) & (.029) & (.075) \tabularnewline
			High education = 1 & .064 & -.015 & -.029 \tabularnewline
			& (.232) & (.769) & (.482) \tabularnewline
			Difference         & -.053 & .078 & .038 \tabularnewline
			& (.441) & (.249) & (.496) \tabularnewline
			\hline
			Italian language = 0 & .145*** & -.045 & -.065* \tabularnewline
			& (.001) & (.364) & (.088) \tabularnewline
			Italian language = 1 & .049 & -.073* & -.036 \tabularnewline
			& (.333) & (.097) & (.374) \tabularnewline
			Difference           & -.097 & -.027 & .030 \tabularnewline
			& (.152) & (.682) & (.592) \tabularnewline
			\hline\hline
		\end{tabular}
	}
	
\vspace{0.4em}
\begin{minipage}{0.82\textwidth}
	\scriptsize
	\setlength{\parskip}{0pt}
	\setlength{\parindent}{0pt}
	\textit{Notes:} The table reports heterogeneous treatment effects by predetermined characteristics. Rows labeled ``=0'' and ``=1'' report average marginal effects for respondents below and above the corresponding threshold. Rows labeled ``Difference'' report the p-value of a test for equality of treatment effects across groups. Columns correspond to different experimental vignettes. P-values are reported in parentheses. * \(p<0.10\), ** \(p<0.05\), *** \(p<0.01\).
\end{minipage}
	
\end{table}

\begin{table}[H]\centering
	\ContinuedFloat
	
	{\footnotesize
		\setlength{\tabcolsep}{5pt}
		\renewcommand{\arraystretch}{0.78}
		
		```
		\textit{Panel B. Job and firm characteristics}
		
		\vspace{0.2em}
		
		\begin{tabular}{lccc}
			\hline\hline
			& Vignette 4 & Vignette 5 & Vignette 7 \tabularnewline
			\hline
			High job tenure = 0 & .058 & .005 & -.010 \tabularnewline
			& (.249) & (.916) & (.790) \tabularnewline
			High job tenure = 1 & .128*** & -.120*** & -.086** \tabularnewline
			& (.007) & (.007) & (.027) \tabularnewline
			Difference & .070 & -.126* & -.076 \tabularnewline
			& (.314) & (.058) & (.169) \tabularnewline
			\hline
			High work experience = 0 & .128** & -.049 & -.038 \tabularnewline
			& (.011) & (.315) & (.313) \tabularnewline
			High work experience = 1 & .059 & -.072 & -.051 \tabularnewline
			& (.211) & (.107) & (.222) \tabularnewline
			Difference & -.070 & -.022 & -.013 \tabularnewline
			& (.315) & (.737) & (.814) \tabularnewline
			\hline
			Public sector = 0 & .051 & -.090** & -.038 \tabularnewline
			& (.227) & (.028) & (.277) \tabularnewline
			Public sector = 1 & .170*** & -.002 & -.071 \tabularnewline
			& (.003) & (.974) & (.132) \tabularnewline
			Difference & .119* & .088 & -.033 \tabularnewline
			& (.100) & (.215) & (.579) \tabularnewline
			\hline
			Large firm = 0 & .084* & -.103** & -.069* \tabularnewline
			& (.075) & (.027) & (.081) \tabularnewline
			Large firm = 1 & .095* & -.012 & -.033 \tabularnewline
			& (.057) & (.798) & (.392) \tabularnewline
			Difference & .012 & .091 & .036 \tabularnewline
			& (.865) & (.163) & (.527) \tabularnewline
			\hline
			Long commuting time = 0 & .068 & -.128** & -.023 \tabularnewline
			& (.187) & (.011) & (.580) \tabularnewline
			Long commuting time = 1 & .112** & -.010 & -.070* \tabularnewline
			& (.013) & (.827) & (.062) \tabularnewline
			Difference & .044 & .118* & -.046 \tabularnewline
			& (.518) & (.075) & (.414) \tabularnewline
			\hline
			Remote work = 0 & .103*** & -.075** & -.057* \tabularnewline
			& (.008) & (.039) & (.078) \tabularnewline
			Remote work = 1 & .066 & -.013 & -.025 \tabularnewline
			& (.363) & (.858) & (.636) \tabularnewline
			Difference & -.038 & .062 & .032 \tabularnewline
			& (.646) & (.449) & (.607) \tabularnewline
			\hline
			Service sector = 0 & .128 & .053 & -.059 \tabularnewline
			& (.137) & (.525) & (.393) \tabularnewline
			Service sector = 1 & .086** & -.084** & -.047 \tabularnewline
			& (.020) & (.019) & (.116) \tabularnewline
			Difference & -.042 & -.137 & .011 \tabularnewline
			& (.654) & (.133) & (.879) \tabularnewline
			\hline\hline
		\end{tabular}
		```
		
	}
	
	\vspace{0.4em}
	\begin{minipage}{0.82\textwidth}
		\scriptsize
		\setlength{\parskip}{0pt}
		\setlength{\parindent}{0pt}
		\textit{Notes:} The table reports heterogeneous treatment effects by predetermined characteristics. Rows labeled `=0'' and `=1'' report average marginal effects for respondents below and above the corresponding threshold. Rows labeled ``Difference'' report the p-value of a test for equality of treatment effects across groups. Columns correspond to different experimental vignettes. P-values are reported in parentheses. * (p<0.10), ** (p<0.05), *** (p<0.01).
	\end{minipage}
	
\end{table}


\begin{table}[H]\centering
	\caption{Heterogeneous treatment effects by age and income groups --- Activism}
	\label{tab:heterogeneity_activism_age_income}
	
	{\footnotesize
		\setlength{\tabcolsep}{5pt}
		\renewcommand{\arraystretch}{0.78}
		
		```
		\begin{tabular}{lccc}
			\hline\hline
			& Vignette 4 & Vignette 5 & Vignette 7 \tabularnewline
			\hline
			\textbf{Age groups} & & & \tabularnewline
			Age group 1 (young) & .088 & -.029 & -.032 \tabularnewline
			& (.179) & (.665) & (.503) \tabularnewline
			Age group 2 (middle) & .117** & -.052 & -.066 \tabularnewline
			& (.022) & (.281) & (.114) \tabularnewline
			Age group 3 (older) & .053 & -.119** & -.024 \tabularnewline
			& (.411) & (.046) & (.693) \tabularnewline
			Age groups (omnibus test) & .605 & 1.180 & .445 \tabularnewline
			& (.739) & (.554) & (.801) \tabularnewline
			\hline
			
			\textbf{Individual income groups} & & & \tabularnewline
			Income group 1 (low) & .093** & -.090** & -.118*** \tabularnewline
			& (.022) & (.021) & (.000) \tabularnewline
			Income group 2 (middle) & .141** & .048 & .115** \tabularnewline
			& (.047) & (.495) & (.027) \tabularnewline
			Income group 3 (high) & -.025 & -.114 & .127 \tabularnewline
			& (.847) & (.342) & (.229) \tabularnewline
			Income groups (omnibus test) & 1.313 & 3.129 & 16.910*** \tabularnewline
			& (.519) & (.209) & (.000) \tabularnewline
			\hline
			
			\textbf{Household income groups} & & & \tabularnewline
			Household income group 1 (low) & .095** & -.085** & -.121*** \tabularnewline
			& (.020) & (.029) & (.000) \tabularnewline
			Household income group 2 (middle) & .144** & .048 & .119** \tabularnewline
			& (.041) & (.502) & (.025) \tabularnewline
			Household income group 3 (high) & -.064 & -.329** & .124 \tabularnewline
			& (.703) & (.038) & (.319) \tabularnewline
			Household income groups (omnibus test) & 1.365 & 5.577* & 16.446*** \tabularnewline
			& (.505) & (.062) & (.000) \tabularnewline
			\hline\hline
		\end{tabular}
		```
		
	}
	
	\vspace{0.4em}
	\begin{minipage}{0.82\textwidth}
		\scriptsize
		\setlength{\parskip}{0pt}
		\setlength{\parindent}{0pt}
		\textit{Notes:} The table reports heterogeneous treatment effects by age, individual income, and household income groups. Age, income, and household income are each divided into three ordered categories. Rows labeled ``omnibus test'' report the p-value of a joint test for equality of treatment effects across all groups within the corresponding block. Columns correspond to different experimental vignettes. P-values are reported in parentheses. * (p<0.10), ** (p<0.05), *** (p<0.01).
	\end{minipage}
	
\end{table}


\clearpage
\section{Figures}
\label{app:figures}

\subsection{Wordclouds}

\begin{figure}[htbp]
\centering
\begin{tabular}{cc}

\subfigure[Control: Amenities]{
\includegraphics[width=0.45\linewidth]{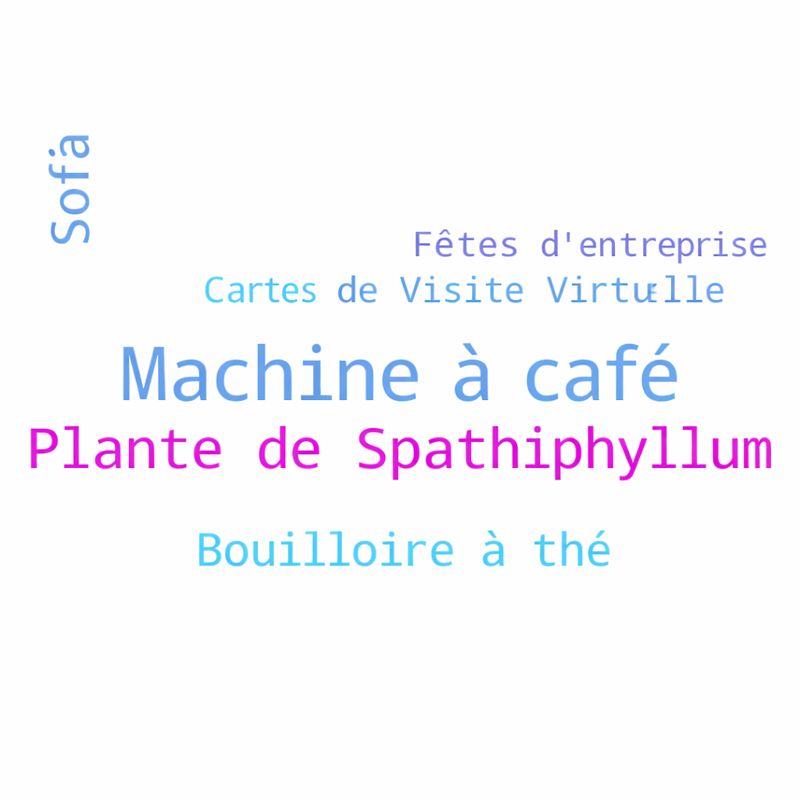}
}
&
\subfigure[T1: Activism]{
\includegraphics[width=0.45\linewidth]{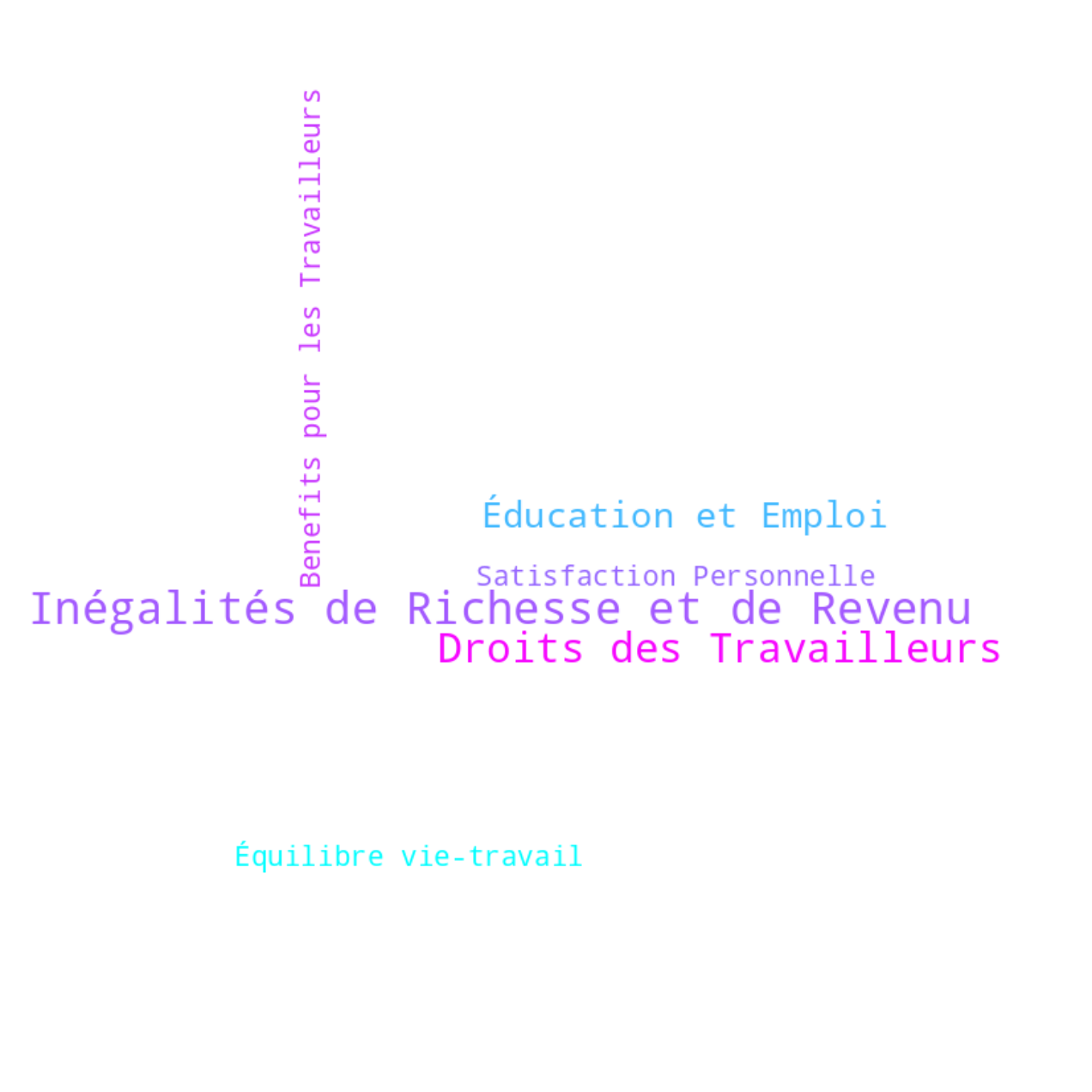}
}
\\[0.4cm]

\subfigure[T2: Work]{
\includegraphics[width=0.45\linewidth]{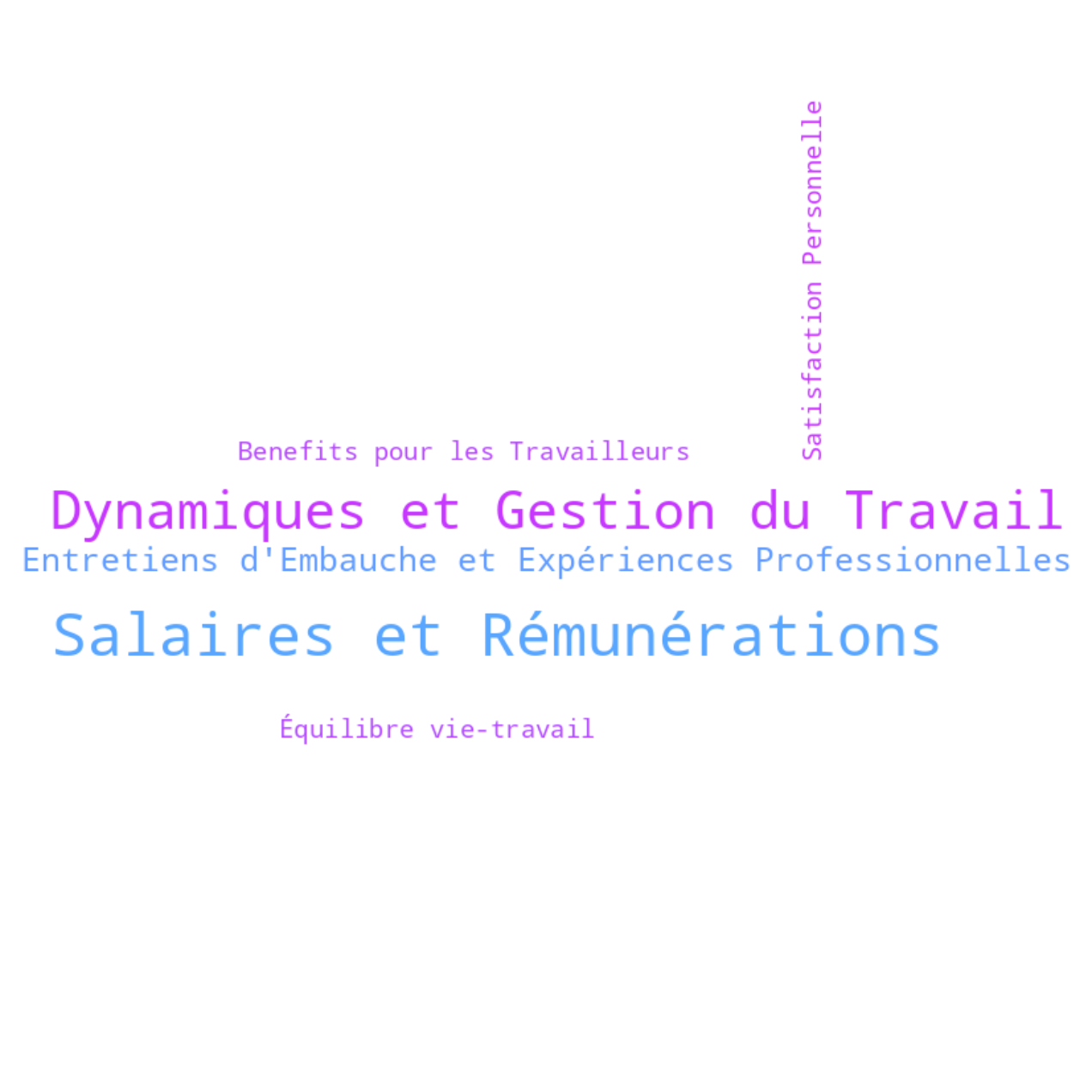}
}
&
\subfigure[T3: Work + Activism]{
\includegraphics[width=0.45\linewidth]{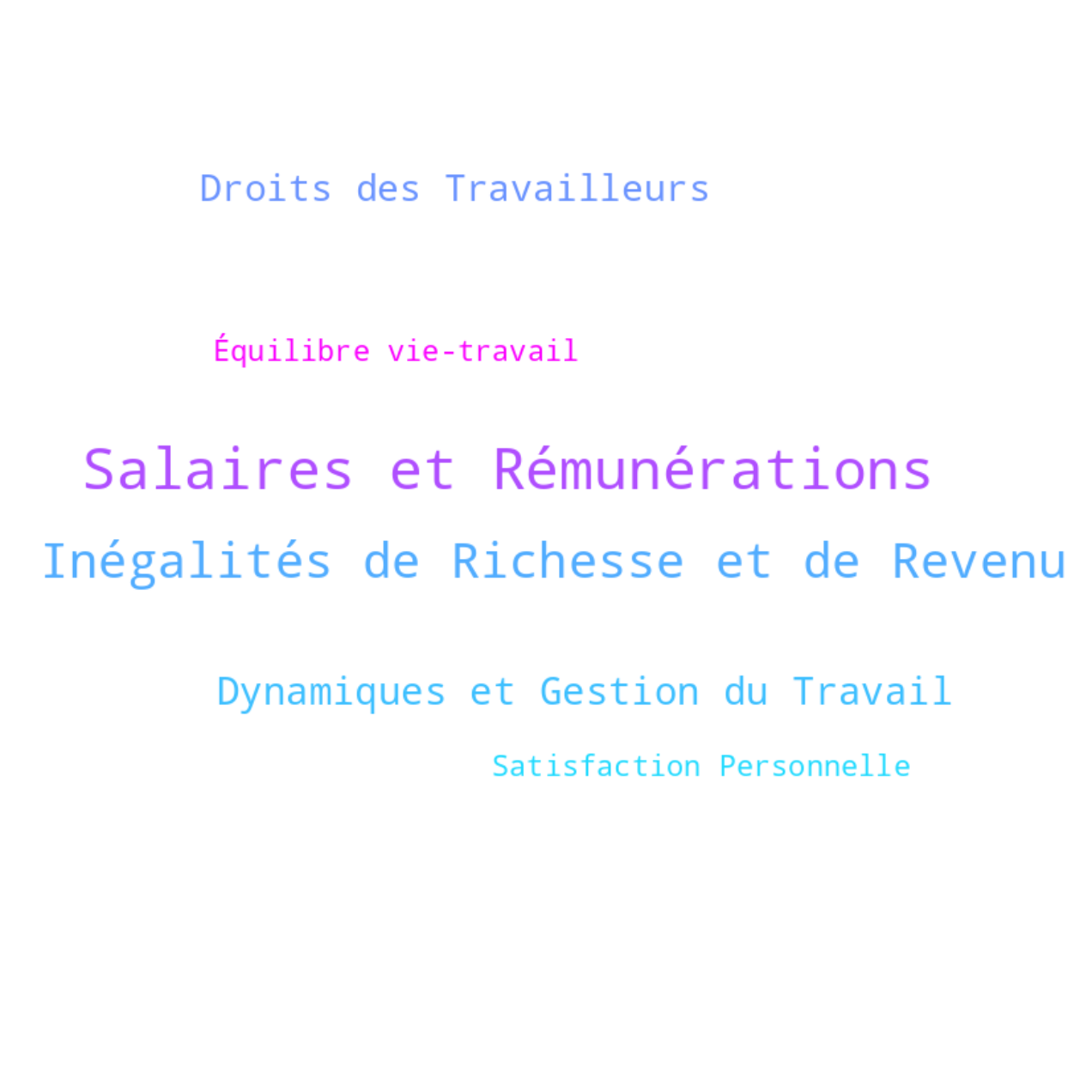}
}

\end{tabular}
\caption{Wordcloud-based experimental conditions (French).}
\label{fig:wordcloud_fra}
\end{figure}
\begin{minipage}{0.99\textwidth}
        \footnotesize
        \textit{Notes:} Wordcloud visualizations used as experimental stimuli (French version). 
Each panel represents one experimental condition. Word size reflects the relative frequency 
of keywords identified through BERTopic analysis of r/antiwork posts (\cite{varavallo2023moral}). 
Health-related terms are included symmetrically across treatments to hold constant the 
Health, Well-being, and Lifestyle dimension.
\end{minipage}

\clearpage
\section{Survey Instrument}
\label{app:survey}

This appendix reports the full questionnaire administered in the survey. The questionnaire was originally administered in Italian and French and is translated here into English.

\subsection*{Introduction}
\begin{quote}
Dear participant,

Thank you for taking part in this study.

Please pay close attention and answer honestly.
Note: it is not possible to return to previous pages to change your answers.

Thank you.

\end{quote}

\subsection{Socio-Demographic Information}

\textbf{Age.} How old are you?
\begin{itemize}
    \item Under 18
    \item 18--24
    \item 25--34
    \item 35--44
    \item 45--54
    \item 55--64
    \item 65 or older
\end{itemize}

\textbf{Gender.} How do you describe yourself?
\begin{itemize}
    \item Male
    \item Female
    \item Non-binary / Third gender
    \item Prefer to self-describe: \underline{\hspace{4cm}}
    \item Prefer not to say
\end{itemize}

\textbf{Marital status.}
\begin{itemize}
    \item Married
    \item Living with a partner
    \item Widowed
    \item Divorced / Separated
    \item Never married
\end{itemize}

\textbf{Country of birth:} \underline{\hspace{6cm}}

\textbf{Postal code of residence:} \underline{\hspace{6cm}}

\textbf{Area of residence.}
\begin{itemize}
    \item City center
    \item Suburban area
    \item Small town
    \item Peripheral area (within one hour by car from a city)
    \item Rural or inner area (more than one hour by car from any city)
\end{itemize}

\textbf{Household size.}
\begin{itemize}
    \item 1
    \item 2
    \item 3
    \item 4
    \item 5 or more
\end{itemize}

\textbf{Number of children under 18 living with you:} \underline{\hspace{3cm}}

\textbf{Highest level of education completed.}
\begin{itemize}
    \item Primary education
    \item Lower secondary education
    \item Upper secondary education (non-university track)
    \item Upper secondary education (university track)
    \item Post-secondary non-university education ($<$ 2 years)
    \item Post-secondary non-university education ($\geq$ 2 years)
    \item Bachelor's degree
    \item Master's degree / Single-cycle degree (4--6 years)
    \item PhD
\end{itemize}

\subsection{Employment History and Job Characteristics}

\textbf{Year of first job:} \underline{\hspace{3cm}}

\textbf{Year of start of current main job:} \underline{\hspace{3cm}}

\textbf{Sector of main job.}
\begin{itemize}
    \item Agriculture, forestry, and fishing
    \item Mining and quarrying
    \item Manufacturing
    \item Electricity, gas, steam, and air conditioning supply
    \item Water supply, sewerage, waste management
    \item Construction
    \item Wholesale and retail trade; repair of motor vehicles
    \item Transportation and storage
    \item Accommodation and food service activities
    \item Information and communication
    \item Financial and insurance activities
    \item Real estate activities
    \item Professional, scientific, and technical activities
    \item Administrative and support service activities
    \item Public administration and defence; compulsory social security
    \item Education
    \item Human health and social work activities
    \item Arts, entertainment, and recreation
    \item Other service activities
    \item Activities of households as employers
    \item Activities of extraterritorial organizations
\end{itemize}

\textbf{Do you work in the public or private sector?}
\begin{itemize}
    \item Public
    \item Private
\end{itemize}

\textbf{Employment status (private sector only).}
\begin{itemize}
    \item Employee
    \item Self-employed
    \item Other: \underline{\hspace{4cm}}
\end{itemize}

\textbf{Number of employees at your workplace.}
\begin{itemize}
    \item 1--10
    \item 11--19
    \item 20--24
    \item 25--49
    \item 50--249
    \item 250--499
    \item 500 or more
\end{itemize}

\textbf{Type of employment contract.}
\begin{itemize}
    \item Full-time
    \item Part-time
    \item Other: \underline{\hspace{4cm}}
\end{itemize}

\textbf{If part-time, reason:}
\begin{itemize}
    \item Financially secure, working by choice
    \item Earn enough working part-time
    \item Want to spend more time with family
    \item Domestic responsibilities
    \item Lack of childcare services
    \item Other: \underline{\hspace{4cm}}
\end{itemize}

\textbf{Working days per week:} \underline{\hspace{3cm}}

\textbf{Working hours per day:} \underline{\hspace{3cm}}

\textbf{Occupational group.}
\begin{itemize}
    \item Legislators, senior officials, and managers
    \item Professionals
    \item Technicians and associate professionals
    \item Clerical support workers
    \item Service and sales workers
    \item Skilled agricultural, craft, and related trades workers
    \item Plant and machine operators, and assemblers
    \item Elementary occupations
    \item Armed forces
\end{itemize}

\textbf{Monthly net wage (12 months):} \underline{\hspace{3cm}}

\textbf{Total household income in the last 12 months.}
\begin{itemize}
    \item Less than \euro 25,000
    \item \euro 25,000--49,999
    \item \euro 50,000--99,999
    \item \euro 100,000--199,999
    \item More than \euro 200,000
    \item Prefer not to say
\end{itemize}

\textbf{Commute time (minutes):} \underline{\hspace{3cm}}

\textbf{Main mode of transportation to work.}
\begin{itemize}
    \item Car / van
    \item Motorcycle / scooter
    \item Bicycle
    \item Bus / coach
    \item Train
    \item Metro / tram
    \item Taxi
    \item Walking
    \item Other: \underline{\hspace{4cm}}
\end{itemize}

\textbf{Do you perform paid remote work (smart working)?}
\begin{itemize}
    \item Yes
    \item No
\end{itemize}

\textbf{If yes, how often do you work from home?}
\begin{itemize}
    \item More than 3 days per week
    \item 3 days per week
    \item 2 days per week
    \item 1 day per week
    \item Occasionally / on request
\end{itemize}

\textbf{If no, could you perform your job from home?}
\begin{itemize}
    \item Completely (100\% or more efficiency)
    \item Mostly (80--90\%)
    \item Partly (50--70\%)
    \item Barely (less than 50\%)
    \item Not at all
\end{itemize}

\textbf{Before COVID-19 (2019), how often did you work remotely?}
\begin{itemize}
    \item Never
    \item Rarely (less than once a month)
    \item Once or twice a month
    \item 1 day per week
    \item 2 days per week
    \item 3 days per week or more
\end{itemize}

\subsection{Experimental Ranking Tasks}

Respondents were shown a word cloud summarizing workplace characteristics mentioned most frequently by a separate sample of workers.

They were asked to rank the following items from most to least important for the sample:

\begin{itemize}
    \item Coffee machine
    \item Spathiphyllum plant
    \item Tea kettle
    \item Digital business cards
    \item Sofa
    \item Company parties
\end{itemize}

Respondents then ranked the same items according to their own personal opinion and experience.

Similar ranking tasks were administered in the treatment conditions, with lists referring to:
(i) work organization and employment conditions,
(ii) social justice and workers' rights,
or (iii) a combination of the two.

\subsection{Outcomes}

\textbf{Agreement statements}  
(1 = Strongly disagree, 4 = Strongly agree)

\begin{itemize}
    \item I give my best at work.
    \item I do not care about my job.
    \item I feel inspired by my job.
    \item I do the minimum amount of work required without going beyond.
    \item I do not express my ideas at work because I fear being assigned more work.
    \item I take as many breaks as possible at work.
    \item I often take initiative at work.
\end{itemize}

\textbf{Frequency of behaviors}  
(1 = Never, 5 = Always)

\begin{itemize}
    \item I help colleagues complete their tasks.
    \item I respond to work-related messages during my days off.
    \item I arrive late at work or leave early.
    \item I call in sick even when I could work.
    \item I pretend to work to avoid receiving another task.
\end{itemize}

\textbf{Probabilities (0--100)}

\begin{itemize}
    \item Leaving your job in the next 12 months.
    \item Seeking help from trade unions.
    \item Reducing your effort at work.
\end{itemize}

\textbf{Suppose that a new law requires employers to offer monetary compensation for the amount of work not performed from home. Which option would you prefer?}

\begin{itemize}
    \item Working from home 5 days per week and keeping your current salary.
    \item Working from home 4 days per week and receiving a 2\% increase in your salary.
    \item Working from home 3 days per week and receiving a 4\% increase in your salary.
    \item Working from home 2 days per week and receiving a 6\% increase in your salary.
    \item Working from home 1 day per week and receiving an 8\% increase in your salary.
    \item Not working from home and receiving a 10\% increase in your salary.
\end{itemize}

\vspace{0.3cm}

\textbf{How much of your net salary would you be willing to give up in order to work one additional day from home?}

\begin{itemize}
    \item 1\% or less of my salary
    \item 2\% of my salary
    \item 4\% of my salary
    \item 6\% of my salary
    \item 8\% of my salary
    \item 10\% of my salary
    \item More than 10\% of my salary
\end{itemize}

\subsection{Workplace Perceptions}

\begin{itemize}
    \item Diversity in hiring practices (Yes/No)
    \item Appreciation of different perspectives (Yes/No)
    \item Sense of community at work (Yes/No)
    \item Fairness of compensation (Yes/No)
    \item Work--life balance (scale 0--100)
    \item Supervisor support (Yes/No)
    \item Organizational support for personal well-being (Yes/No)
    \item Flexibility of working hours (Yes/No)
    \item Ability to use vacation days (Yes/No)
    \item Time for family and friends (Yes/No)
    \item Reduction of hobbies due to work (Yes/No)
    \item Resources to manage work-related stress (Yes/No)
    \item Experience of burnout (Yes/No)
    \item Experience of toxic positivity (Never--Always)
    \item Perceived impact of toxic positivity on well-being (Not at all--Extremely)
    \item Willingness to recommend current employer (Yes/No)
\end{itemize}

\subsection*{Final Comment}

\textbf{Optional comment:} \underline{\hspace{10cm}}


\end{document}